\documentclass[final,5p,times,twocolumn]{elsarticle}

\usepackage[utf8]{inputenc}
\usepackage[numbers]{natbib}
\usepackage{pdflscape}

\usepackage{float}

\usepackage{graphicx}
\usepackage{tabularx}
\usepackage{adjustbox}
\usepackage{tikz}
\usepackage{forest}
\usepackage{titlesec}
\usepackage{longtable}
\usepackage{makecell}
\usepackage{amssymb}
\usepackage{amsmath} 
\usepackage{multirow} 
\usepackage{xcolor}
\usepackage{enumitem}
\usepackage{svg}
\usepackage{booktabs} 
\usepackage[colorlinks=true, linkcolor=black, citecolor=black, urlcolor=black]{hyperref}

\definecolor{darkgreen}{RGB}{0, 100, 0}  
\definecolor{lightviolet}{RGB}{230, 230, 250} 
\definecolor{violet}{RGB}{199, 184, 230} 
\definecolor{darkviolet}{RGB}{226,183,226} 

\usepackage{color}
\titleformat*{\section}{\normalsize\bfseries}
\titleclass{\subsubsubsection}{straight}[\subsection]
\newcounter{subsubsubsection}[subsubsection]
\renewcommand{\thesubsubsubsection}{\thesubsubsection.\arabic{subsubsubsection}}
\titleformat{\subsubsubsection}
 {\normalfont\normalsize\itshape}{\thesubsubsubsection}{1em}{}
\titlespacing*{\subsubsubsection}
{0pt}{3.25ex plus 1ex minus .2ex}{1.5ex plus .2ex}

\usetikzlibrary{shapes,arrows.meta,positioning}
\tikzstyle{block} = [rectangle, draw, fill=blue!20, text width=5em, text centered, rounded corners, minimum height=4em]
\tikzstyle{line} = [draw, -Latex]
\def\tsc#1{\csdef{#1}{\textsc{\lowercase{#1}}\xspace}}
\tsc{WGM}
\tsc{QE}
\tsc{EP}
\tsc{PMS}
\tsc{BEC}
\tsc{DE}

\date{}
\journal{arXiv}

\begin{document}
\let\WriteBookmarks\relax
\def\floatpagepagefraction{1}
\def\textpagefraction{.001}

\begin{frontmatter}
\title{A Systematic Review of Generalization Research in Medical Image Classification}
\author[1,2]{Sarah Matta}
\ead{sarah.matta@univ-brest.fr}
\address[1]{Université de Bretagne Occidentale, Brest, Bretagne, 29200 France}

\address[2]{Inserm, UMR 1101, Brest, F-29200, France}
\author[1,2]{Mathieu Lamard}
\author[3,2,1]{Philippe Zhang}
\author[3]{Alexandre Le Guilcher}
\author[3]{Laurent Borderie}
\address[3]{Evolucare Technologies, Villers-Bretonneux, F-80800, France}
\author[1,2,4]{Béatrice Cochener}
\address[4]{Service d’Ophtalmologie, CHRU Brest, Brest, F-29200, France}
\author[2]{Gwenolé Quellec}

\begin{abstract}
Numerous Deep Learning (DL) classification models have been developed for a large spectrum of medical image analysis applications, which promises to reshape various facets of medical practice. Despite early advances in DL model validation and implementation, which encourage healthcare institutions to adopt them, a fundamental questions remain: how can these models effectively handle domain shift? This question is crucial to limit DL models performance degradation. Medical data are dynamic and prone to domain shift, due to multiple factors. Two main shift types can occur over time: 1) covariate shift mainly arising due to updates to medical equipment and 2) concept shift caused by inter-grader variability. To mitigate the problem of domain shift, existing surveys mainly focus on domain adaptation techniques, with an emphasis on covariate shift. More generally, no work has reviewed the state-of-the-art solutions while focusing on the shift types. This paper aims to explore existing domain generalization methods for DL-based classification models through a systematic review of literature. It proposes a taxonomy based on the shift type they aim to solve. Papers were searched and gathered on Scopus till 10 April 2023, and after the eligibility screening and quality evaluation, 77 articles were identified. Exclusion criteria included:  lack of methodological novelty (e.g., reviews, benchmarks), experiments conducted on a single mono-center dataset, or articles not written in English. The results of this paper show that learning based methods are emerging, for both shift types. Finally, we discuss future challenges, including the need for improved evaluation protocols and benchmarks, and envisioned future developments to achieve robust, generalized models for medical image classification.
\end{abstract}


\onecolumn


\end{frontmatter}

\section{Introduction}
\label{sec:Introduction}
Deep Learning (DL) models are the current state-of-the-art method for medical image classification. The availability of high quality labeled data, typically through multi-site collaboration projects, has paved the way to employ these data driven-based approaches in supervised medical image analysis. Nowadays, DL models have achieved human level performances in different medical domains such as dermatology \citep{esteva2017dermatologist}, oncology \citep{luo2019real}, histopathology \citep{hegde2019similar} and ophthalmology \citep{de2018clinically}. 

Current large-scale clinical DL models are often trained using a single large dataset collected from a specific population, typically through a partnership with one healthcare institution. Once the models have been approved by regulatory authorities, they should be deployed to different populations, image acquisition protocols or devices. In such cases, it is important to ensure that the performance drop is minimal. However, recent prospective validation studies have shown significant decreases in model performance when confronted to domain shifts across different institutions, notably in the contexts of chest X-rays \citep{cohen2020limits, pooch2020can, zech2018variable}, MRIs \citep{albadawy2018deep, maartensson2020reliability}, pathology \citep{stacke2020measuring, stacke2019closer, thagaard2020can} and fundus photography \citep{matta2023towards}. This is mainly because the assumption that training and testing data are drawn from the same distribution (Independent and Identically Distributed (IID) assumption) for which most of the DL models rely on, may be not hold in real-world scenarios. 

More generally, the differences between the training and testing data are defined as shifts between the respective data distributions. These data distributions can be expressed as the product of the probability of the input data $p(x)$ and the conditional probability of the output labels given the input data $p(y|x)$, resulting in the joint distribution $p(x,y)$. The IID setup, also known as within-distribution generalization, corresponds to the traditional evaluation form where there is no shift in data distributions, $p(x_{testing})=p(x_{training})$ and $p(y_{testing}|x_{testing}) =p(y_{training}|x_{training})$. This type of evaluation is the simplest form of generalization. The more challenging setup, the non-IID setup, corresponds to the other cases where shift occur between train and test data distribution. These cases are commonly referred to as out-of-distribution (OOD) shifts \citep{hupkes2023taxonomy}. 

While characterization of this OOD shift is still an open problem, recent work \citet{cohen2020limits, shen2021towards} have identified two main data shift types: \textit{the covariate shift} and \textit{the concept shift}. The \textit{covariate shift}, the most commonly considered data distribution shift in OOD, occurs when the distribution of the data changes $ p(x_{testing})  \neq p(x_{training})$, while keeping the conditional probability of the labels given the input  $p(y_{testing}|x_{testing}) =p(y_{training}|x_{training})$ (which describes the task). On the other hand, the \textit{concept shift} corresponds to the case where the relationship between the input and class variables changes \citep{moreno2012unifying}. In other terms, $p(y_{testing}|x_{testing}) \neq p(y_{training}|x_{training})$. 

In practice, in the medical field, covariate shift can occur due to the data heterogeneity caused by using different acquisition protocols across medical centers (difference in staining procedure, multi-vendor scanners/cameras, variable acquisition parameters) which might lead to variability in terms of illumination, color or optical artifacts. Moreover, obtaining high quality image is not always guaranteed, and images may be low quality due to using low-cost imaging systems or due to tissue preparation or preservation artifacts. In some cases, it can also be prone to the operator subjectivity such as in ultrasound or endoscopy imaging, where the operator moves the device.

On the other hand, concept shift is mainly caused by label noise. In fact, the challenge reside in collecting accurate labeled medical image dataset. Manual annotations are error-prone, tedious, and time-consuming. In addition, as labels are provided by experts, certain level of subjectivity is expected. In fact, different classification systems for disease may be adopted. For instance, for Diabetic Retinopathy (DR) screening grading and management, different disease severity scales exist, such as the International Classification for Diabetic Retinopathy (ICDR), the English DR NHS, the Scottish DR grading scheme, the Canadian Tele-Screening Grading \citep{boucher2020evidence}, and the French DR grading which follows the International Grading System \citep{wilkinson2003proposed}.

Generalizing DL models is considered to be one of the biggest challenges facing a wider adoption and successful deployment of DL models in medical applications.
To cope with this serious problem, recent effort has focused on improving DL model generalizability and developing robust DL models in non-IID settings. 
A straightforward solution to mitigate data heterogeneity and this distribution shift problem in medical imaging is to adapt DL models to the target domain using \textit{Domain Adaptation (DA)} methods. DA methods can be categorized into \textit{Supervised Domain Adaptation (SDA)} and \textit{Unsupervised Domain Adaptation (UDA)} techniques based on the availability of labels in the target domain. In SDA, a limited amount of labeled data from the test domain is available for training the DL models. Typically, this involves \textit{transfer learning}, where a pre-trained DL model on a large dataset from the source domain is fine-tuned on the targeted dataset using supervised learning.
In contrast, UDA methods focus on scenarios where labeled data in the target domain is not available and only unlabeled target data are available for training. It aims to transfer the knowledge from a label-rich training (e.g source) domain to a test (target) domain, without the need of a labeled target domain. 

However, UDA methods are limited in practice, as they still require access to a part of the test-domain data during the training procedure. To overcome this limitation,  \textit{Domain Generalization} (DG) methods have emerged as a more promising solution. In DG, the goal is to develop a DL model that is able to generalize to one unseen target domain via learning from a single or multiple source domains, without having access to the testing data from the target domain. However, training DG methods using \textit{multi-source data} (multi-DG) has been considered as costly since collecting medical data from multiple sources is challenging, and medical data are subject to privacy regulations. To address this problem, recent work focused on an additional research line, called \textit{single domain generalization} (single-DG), in which the goal is to develop a DL model that is able to generalize to multiple target domains via learning from a single source domain \citep{li2022single}. Alternatively, \textit{semi-supervised domain generalization} \citep{zhang2022semi} combines the single-DG and multi-DG by using one labeled sources domain and several unlabeled source domain to boost the performances.

\section{Aims and scope of this paper}
DG in computer vision dataset is becoming an emerging field: numerous surveys have been proposed \citep{zhou2022domain,wang2022generalizing}. In the medical field, research has focused on domain adaptation \citep{guan2021domain} or unsupervised domain adaptation \citep{kumari2023deep}. Other medical research has reviewed the problem of learning with noisy labels \citep{karimi2020deep,zhou2021review,rathod2021automatic}. However, to the best of our knowledge, no medical review has studied the problem of generalization of DL models in the medical field with a focus on both domain shift problems: covariate shift and concept shift. A study of the current DL methods tackling these problems is thus necessary for guiding practitioners and researchers in understanding the challenges and existing trends in the field. In particular, exploring and analyzing these methods would help identify the limits and the best methods. This would lead to more efficient and robust DL systems, enabling a broader applicability of AI in different environment healthcare settings.
This paper presents the first systematic review of generalization research in medical image classification. It aims to answer the following Research Questions (RQ):
\begin{itemize}
    \item \textbf{RQ1:} What are the state-of-the-art methods in medical image classification targeting domain shift in the literature? \\
    \textbf{Significance:} A taxonomy and a clustering of similar methods would help analyze the performances for different shift types. It would also help identifying which generalization techniques are most effective under different circumstances.
       \item \textbf{RQ2:} What are the related areas in which generalization research can be applied? \\
    \textbf{Significance:} Identifying related areas will help understand the scenarios where this research can be combined with other studies or applied in practice.
    \item \textbf{RQ3:} What are the best practices for implementing generalization techniques in research? \\
  \textbf{Significance:} Identifying open-source libraries and implementations details would enhance generalization research in the medical domain.
    \item \textbf{RQ4:} What are the key challenges and future promises for generalization research?\\ 
    \textbf{Significance:} Identifying key challenges and future research areas with potential for significant advancements in generalization research is crucial for guiding researchers toward the most promising directions.
\end{itemize}

In this paper, we make the following key contributions:
\begin{itemize}
    \item We present the first systematic survey on generalization research for medical image analysis based on covariate shift and concept shift. Based on the assumed shift type, the reader can refer to methods in our taxonomy.
    \item We present public medical datasets and open-source libraries to enhance future research in this field.
    \item We study the recent trends in generalization research and found that learning-based methods are showing an increased interest. In particular, foundation models hold promises for enhanced generalizability.
    \item Our analysis shows that this research is applied to wide areas in medical imaging, including: X-ray, fundus photography, dermoscopic imaging, and pathology. There is a need for benchmarking strategies to better assess these methods.
\end{itemize}

The organization of this paper is as follows. In Section~\ref{sec:sec_domain generalization}, we briefly describe the problem of domain generalization. In Section~\ref{sec: Methodology}, we introduce our methodology for literature review. In Section~\ref{sec:taxonomy_generalization}, we present our taxonomy, in which we review DL methods that have dealt with covariate shift in the medical domain (Section~\ref{sec:covariate_shift}) and DL methods aiming to overcome the problem of concept shift and noisy labels (Section~\ref{sec:concept_shift}). In Section~\ref{sec: Public medical datasets}, we present public medical datasets used for generalization research. Section~\ref{sec: Discussion} discusses the benefit of current DG methods based on the results of challenge data. Furthermore, it presents trends in DG development, related research to DG, implementation details, and future directions. Finally, Section~\ref{sec: Conclusion} concludes this work.

\section {Domain generalization problem formulation}
\label{sec:sec_domain generalization}

Consider $\mathcal{X \times Y}$ as the combined space of images ($\mathcal{X}$) and their respective class labels ($\mathcal{Y}$). Let $\mathcal{S}$ denote the source domain, composed of data sampled from a distribution, $\mathcal{S}=\{(x_j, y_j)\}_{j=1}^{n} \sim p(\mathcal{{X,Y}})$,  where $x_i \in \mathcal{X} \in \mathbb{R}^d$ denotes the sample in the input space, 
$y_i \in \mathcal{Y} \in \mathbb{R}$ designates the label belonging to the output space, 
$n$ is the data size of source domain, 
${p(\mathcal{{X,Y}})}$ is the joint space of images $\mathcal{X}$ and their respective class labels $\mathcal{Y}$.
In domain generalization, $M$ source domains $\mathcal{S}^i=\{(x_j^i, y_j^i)\}_{j=1}^{n_i}$ (where $\mathcal{S}^i$ denotes the $i$-th domain, and $n_i$ is the data size of source domain $i$) are provided for training: $\mathcal{S}_{train}=\{\mathcal{S}^i\ |\ i = 1, ..., M\}$.

DG approaches aim to learn a robust and generalizable predictive function $f: \mathcal{X} \rightarrow \mathcal{Y}$ using the $M$ training source domains and optimizing it to achieve a minimum prediction error on an unseen target domain 
$\mathcal{T} \sim q({\mathcal{X,Y})}$. In contrast to domain adaptation approaches, the target domain is inaccessible during training and is sampled from an unknown and different distribution than the $M$ source domains, that is $p({\mathcal{X,Y}})^{i}\neq q({\mathcal{X,Y}})$ for $i \in \{ 1, ..., M\}$ \citep{wang2022generalizing}. Therefore, the DG objective can be formulated as follows:
\begin{equation}
\min_f \mathop{\mathbb{E}}_ {(x,y)\in \mathcal{T}} [\mathcal{L}(f(x),y)]
\end{equation}
where $\mathop{\mathbb{E}}$ is the expectation and $\mathcal {L(.,.)}$ is the classification loss function.

\section{Research Methodology}
\label{sec: Methodology}

\begin{figure}[h!]
    \centering
\includegraphics[width=8cm,height=10cm]{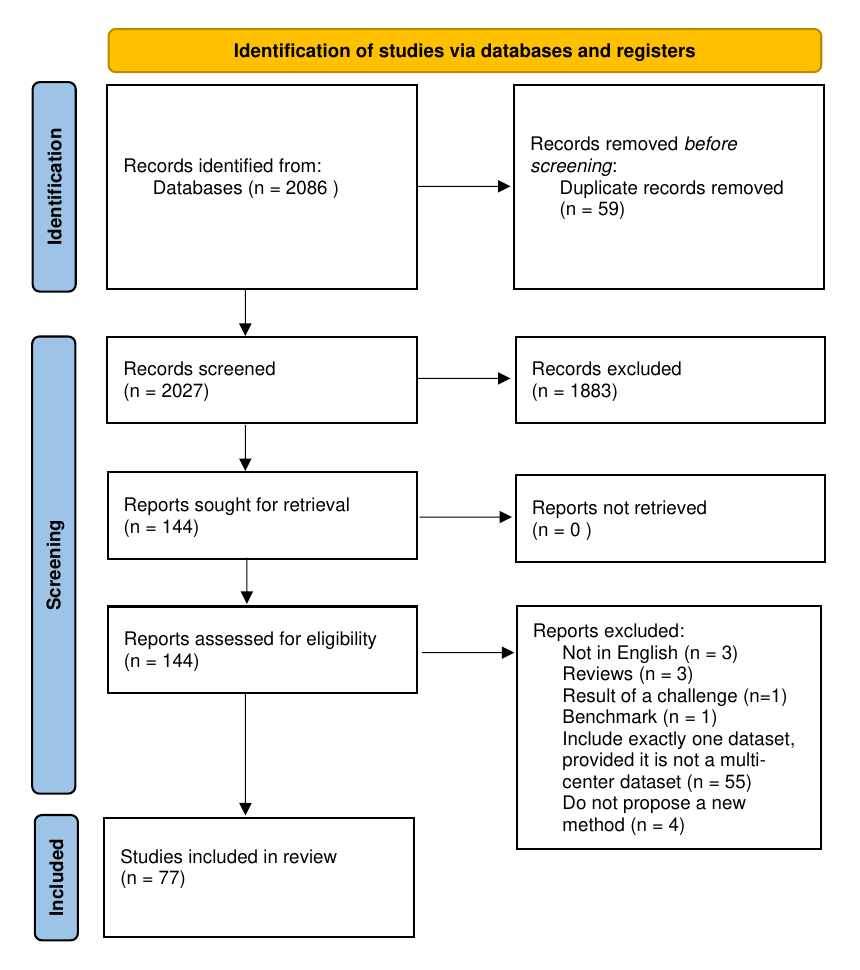} 
\caption{PRISMA flowchart for systematic review of generalizing methods.}
    \label{Fig:Figure_PRISMA}
\end{figure}

To address our research questions posed, the Preferred Reporting Items for Systematic Reviews and Meta-Analyses (PRISMA) guiding principles for conducting systematic reviews \citep{page2021prisma} was applied to select papers which develop solutions in generalization research.
Figure~\ref{Fig:Figure_PRISMA} displays the PRISMA flowchart conducted in this work. 
The review process consisted of gathering studies using Scopus database. The search strategy was piloted by one reviewer using the following query: \lq\lq \textit{domain generalization}\rq\rq\  OR \lq\lq \textit{noisy labels}\rq\rq\ OR  \lq\lq \textit{covariate shift}\rq\rq \ OR \lq\lq \textit{concept shift}\rq\rq. This search was done within Article title, Abstract, and keywords.
We included papers published from 1 January 2020 to 10 April 2023 (included). A total of 2086 papers were found.
First, the search results were reviewed and duplicated records were removed using Zotero.
This resulted in 2027 papers. Abstracts and titles were manually reviewed. Papers were included if they were dealing with medical image classification and deep learning methods. They were excluded if they met one of the following criteria: 1) the paper was not accessible in English, 2) the paper was a review,  3) the paper was a result of a challenge, 4) the paper was a benchmark, 5) the paper included exactly  one dataset provided it is not a multi-center dataset, 6) the paper did not propose a new method for tackling concept shift or covariate shift. When in doubt about the eligibility of the study, the full text was retrieved ant reviewed. 
The total number of papers considered in this survey was 77 papers.

We developed a data extraction form comprising different items related to the research questions. It included the following items: 1) title of the article, 2) year of publishing, 3) modality, 4) organ, 5) task, 6) dataset, 7) type of shift: covariate or concept shift, 8) deep model: the deep learning technique used in the study, 9) code availability and 10) dataset availability. One reviewer collected data from each report.

For a fair comparison, we have chosen to report the results of papers using the same testing subset.

\section{What are the state-of-the-art methods in medical image classification targeting domain shift in the literature?}
\label{sec:taxonomy_generalization}
\begin{figure*}[htbp]
    \centering
\includegraphics[width=19cm,height=22cm]{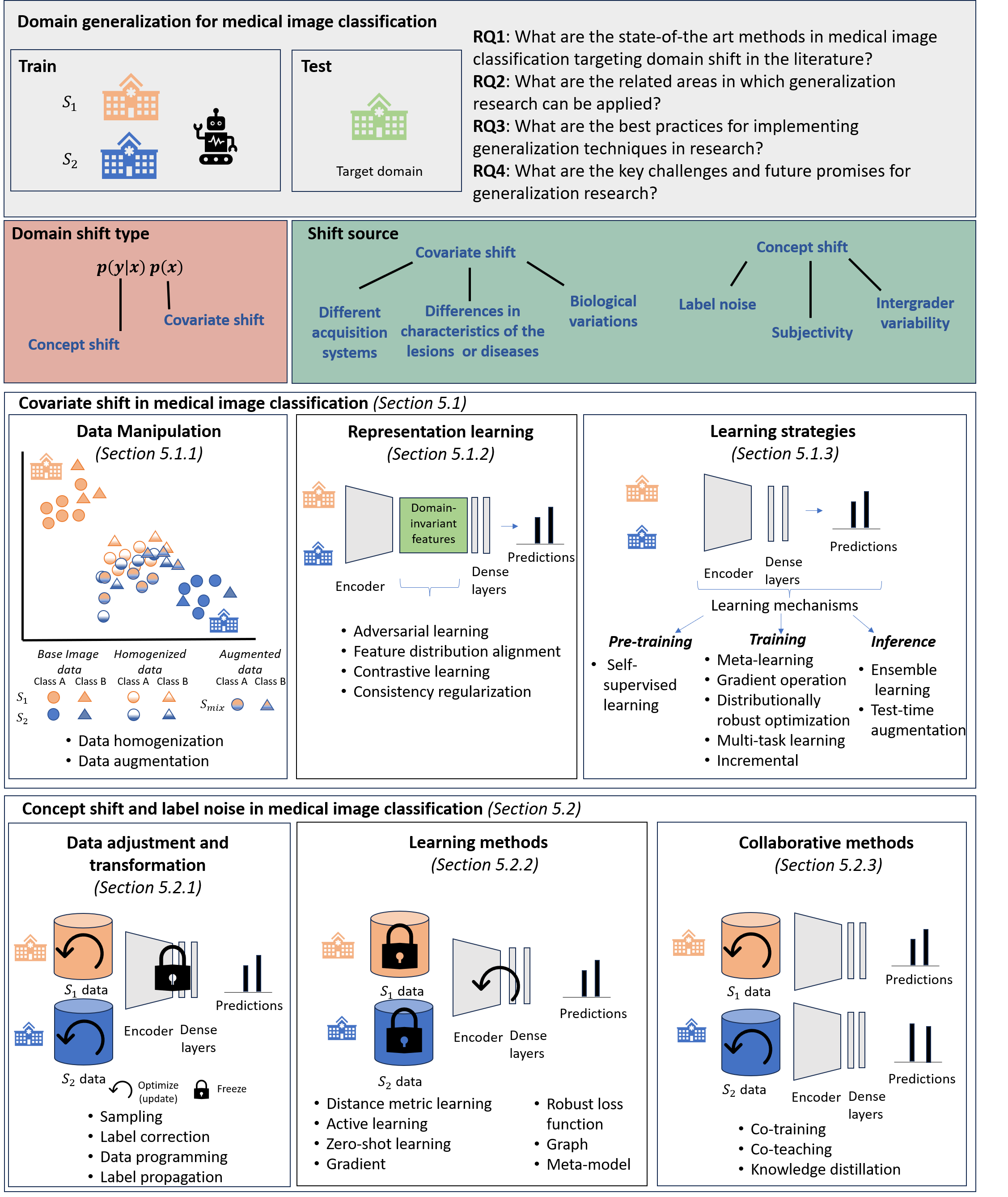} 
\caption{The generalization taxonomy proposed in our domain generalization research for medical image classification. The motivation of this work is based on four research questions. Depending on the domain shift type, 3 methods of categories  were identified for covariate shift and concept shift.
}
    \label{fig:Figure_2}
\end{figure*}

\begin{figure*}[t]
  \centering
    \scalebox{0.7}{
\tikzset{
    basic/.style  = {draw, text width=5cm, rounded corners=2pt, align=center, font=\sffamily\large, rectangle,fill=lightviolet},
    root/.style   = {basic, rounded corners=2pt, thin, align=center, fill=gray!30, font=\sffamily\large},
    onode/.style = {basic, thin, rounded corners=2pt, align=center, fill=green!60,text width=3cm, font=\sffamily\large},
    tnode/.style = {basic, thin, align=left, fill=pink!60, text width=15em, align=center,  text width=40em, font=\sffamily\large},
    xnode/.style = {basic, thin, rounded corners=2pt, align=center, fill=blue!20,text width=8cm, font=\sffamily\large},
    edge from parent/.style={draw=black, edge from parent fork right,font=\sffamily\large}
}
\begin{forest} for tree={
    grow'=east,
    growth parent anchor=west,
    parent anchor=east,
    child anchor=west,
    font=\large,
    edge path={\noexpand\path[\forestoption{edge},->, >={latex}] 
         (!u.parent anchor) -- +(10pt,0pt) |-  (.child anchor) 
         \forestoption{edge label};}
}
[Concept shift and label noise in medical image classification (Section~\ref{sec:concept_shift}), basic, l sep=10mm,
[Data adjustment and transformation (Section~\ref{sec:Data adjustment and transformation}), root,  l sep=10mm,
   [Sampling \citep{son2021leveraging} \citep{xue2022image}
 \citep{aljuhani2022uncertainty}  \citep{bai2021cnngeno} \citep{xu2022meta} \citep{hu2022multi}    (Section~\ref{sec:Sampling}), xnode,  l sep=10mm, 
    ]
       [Label correction \citep{hermoza2022censor} \citep{bai2021novel} \citep{qiu2023hierarchical} 
 \citep{he2022reducing}  \citep{zhu2021hard} \citep{zhu2023robust} (Section~\ref{sec: Re-labeling}), xnode,  l sep=10mm, 
        ]
         [Data programming  \citep{dunnmon2020cross} (Section~\ref{sec:Data programming}), xnode,  l sep=10mm, ]
 [Label propagation \citep{vindas2022semi} \citep{ying2023covid} (Section \ref{Sec:Label propagation}), xnode,  l sep=10mm, 
] ]
 [Learning methods \\ (Section~\ref{sec:Learning methods}), root,  l sep=10mm,
          [Distance metric learning \citep{zhang2022re} \citep{van2022differential} 
          \citep{seibold2022breaking} \citep{kurian2022improved} (Section~\ref{sec:Distance metric learning}), xnode,  l sep=10mm,   
     ]   
        [ Active learning  \citep{son2021leveraging}(Section~\ref{sec:Active learning}), , xnode,  l sep=10mm,
    ]
      [Zero-shot learning  \citep{paul2021generalized} (Section~\ref{Zero-shot learning}), xnode,  l sep=10mm,   
]
[ Gradient \citep{elbatel2022seamless} (Section~\ref{sec: gradient}), , xnode,  l sep=10mm,]
   [Robust loss function \citep{hermoza2022censor}  \citep{qiu2023hierarchical} \citep{zhu2021hard} \citep{sun2022fully} \citep{hu2021deep}  \citep{yu2022space}  \citep{liu2022nvum} \citep{shi2020graph}  
 \citep{gundel2021robust} 
      (Section~\ref{Sec:Robust loss function}), xnode,  l sep=10mm,  ]
       [Graph \citep{ying2023covid} \citep{yu2022space}   \citep{shi2020graph} \citep{xiang2023automatic}  (Section~\ref{sec:Graph}), xnode,  l sep=10mm, 
    ]
     [Meta-model \citep{do2021multiple} (Section~\ref{Meta-model}), xnode,  l sep=10mm, ]
]
       [Collaborative \\ methods (Section~\ref{Sec:Collaborative and iterative methods}), root,  l sep=10mm, 
         [Co-training  \citep{zhou2023combating} \citep{xue2022robust}  (Section \ref{Sec:Co-training}), xnode,  l sep=10mm,  
  ]
       [Co-teaching \citep{hermoza2022censor}   \citep{zhu2021hard} \citep{zhu2023robust}  \citep{liu2021co} \citep{peng2020noise}(Section \ref{Sec:Co-teaching}), xnode,  l sep=10mm, 
   ]
    [Knowledge distillation \citep{li2021bootstrap} (Section \ref{sec:Knowledge distillation}), xnode,  l sep=10mm,   ]]  
  ]
\end{forest}}
    \caption{Literature survey tree for concept shift and label noise.}
    \label{fig:Figure_4}
\end{figure*}
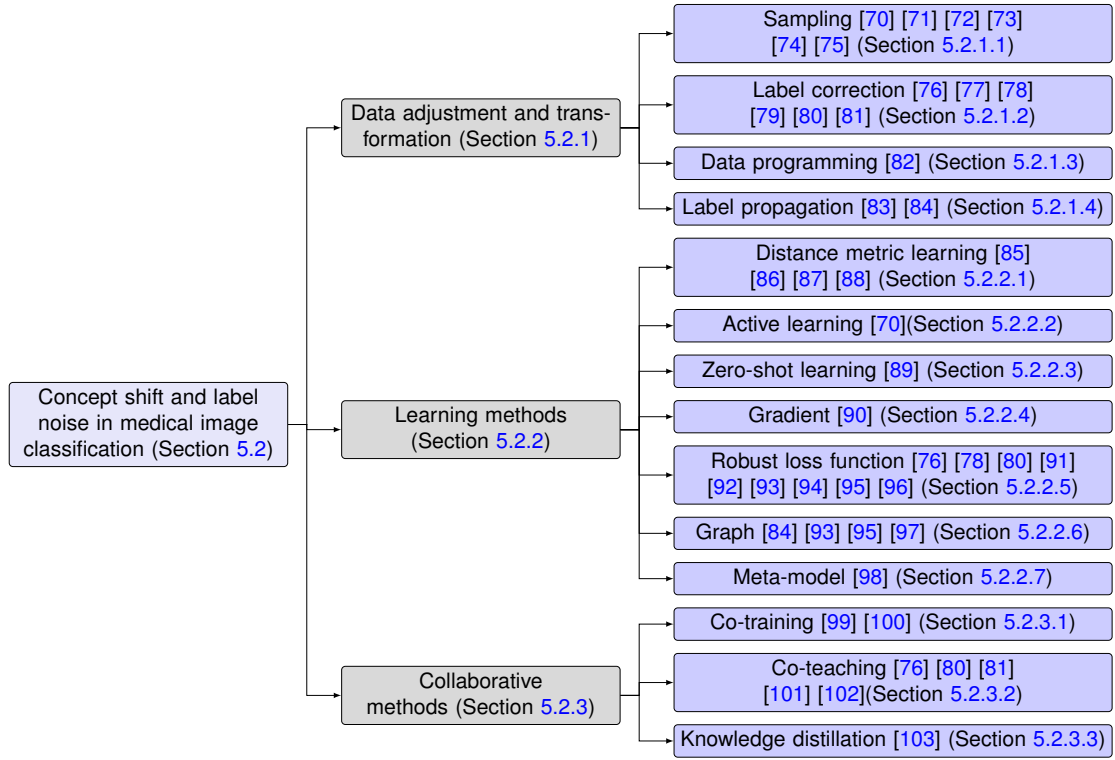

The identified papers were reviewed and a taxonomy was proposed based on the common methodology and the assumed shift they solve.
Depending on the assumed domain shift (covariate shift or concept shift), a plethora of methods have been proposed. To make it easier for the reader to find the methods suited to their problem, we have therefore chosen to first separate the methods based on this criterion (Figure~\ref{fig:Figure_2}). In this section, we present our categorization of methods based on covariate shift (Figure~\ref{fig:Figure_3}) and concept shift (Figure~\ref{fig:Figure_4}). These methods are detailed in the following sections (Section \ref{sec:covariate_shift} and Section \ref{sec:concept_shift}).
Table~\ref{table:Notations} presents the notations used in this paper.  

\begin{table*}[h!]
\centering
\begin{tabular}{ccccc}
\toprule
Notation & Description& &Notation & Description\\
\toprule
$x,y$  & Instance/clean label  &&KL& Kullback-Leibler divergence \\
$\mathcal{X},\mathcal{Y}$ & Feature/label space && $D_{KL}$ & Symmetrized Kullback-Leibler   \\
$Y$ & Pair labels &&$s$& Soft label distribution \\
$\theta$  & Model parameter && $f$ & Network prediction with input $x$ \\
$\mathcal{L(\cdot,\cdot)}$  & Loss function &&$c$& Number of classes\\
$L(\cdot)$ & Cross-entropy loss  &&$e$ & Training epoch \\
$E, C$ & Feature extractor (encoder)/classifier &&$I$ & Individual regularization \\
$M$& Number of source domains  &&$z^c$& Label prediction \\
$C_D$ & Domain classifier (domain discriminator) && $z^f$& Feature representation   \\
$C_S$& Category classifier&&$\tilde{z}$ & Temporal ensembling momentum \\
$f$  & Predictive function &&$m$ & Momentum coefficient \\
$\mathop{\mathbb{E}}$ & Expectation &&$\mu$& Mean value \\
$\mathcal{S}$& Source domain &&$v$, $\delta$, $\beta$, $\gamma$& Hyperparameters \\
$\mathcal{T}$& Target domain &&$w$& Weight \\
$\alpha$& Learnable parameter& &$B$ & Batch of selected images \\
$P$&Total number of positive samples &&$N$ &  Total number of negative samples \\
$p$, $q$ & Distribution&&R & Risk function  \\
$n_i$ & Data size of source domain $i$ &&$n$ & Data size of total training data  \\
$\hat{x}$ & Augmented instance &&$y^d$ & Label distribution  \\
$\hat{y}$ & Noisy label&& $P$ & Total number of positive
samples \\
$K$& Constant &&$\lambda$ & weight parameter \\
TP & True positive&& TN & True negative \\
cov & Covariance &&$\hat{cov}$ & Mean covariance matrix \\
$T$ & Task &&$D$ & Dataset \\
$D^{tr}$ & Training dataset &&$D^{val}$ & Validation dataset \\ 
$D^{test}$ & Testing dataset && $d$ & Distance  \\ 
$\tau$ & Temperature &&  &  \\
\bottomrule
\end{tabular}
\caption{Notations}
\label{table:Notations}
\end{table*}

\begin{figure*}[t]
  \centering
    \scalebox{0.7}{
\tikzset{
    basic/.style  = {draw, text width=5 cm, rounded corners=2pt, align=center, font=\sffamily\large, rectangle,fill= lightviolet},
    root/.style   = {basic, rounded corners=2pt, thin, align=center, fill=gray!30, font=\sffamily\large},
    onode/.style = {basic, thin, rounded corners=2pt, align=center, fill=green!60,text width=3cm,font=\sffamily\large},
    tnode/.style = {basic, thin, align=left, fill=pink!60, text width=15em, align=center,  text width=40em,font=\sffamily\large},
    xnode/.style = {basic, thin, rounded corners=2pt, align=center, fill=blue!20,text width=8cm,font=\sffamily\large},
    edge from parent/.style={draw=black, edge from parent fork right,font=\sffamily\large}
}
\begin{forest} for tree={
    grow'=east,
    growth parent anchor=west,
    parent anchor=east,
    child anchor=west,
    font=\medium,
    edge path={\noexpand\path[\forestoption{edge},->, >={latex}] 
         (!u.parent anchor) -- +(10pt,0pt) |-  (.child anchor) 
         \forestoption{edge label};}
}
[Covariate shift in medical image \\ classification (Section~\ref{sec:covariate_shift}), basic,  l sep=10mm,
    [Data manipulation (Section~\ref{sec:Data manipulation}), root,  l sep=10mm,
        [Data homogenization \citep{kurian2022domain} \citep{yin2022afa} \citep{gunasinghe2022domain} \citep{garrucho2022domain} 
  \citep{wang2021harmonization}  (Section~\ref{sec:Data Homogenization}), xnode,  l sep=10mm,
     ]
    [Data augmentation   \citep{li2022single} \citep{zhang2022semi}  \citep{garrucho2022domain} 
 \citep{lucieri2022revisiting}   \citep{wang2023domain} \citep{lafarge2021rotation} \citep{dexl2021mitodet} \citep{long2021domain} \citep{li2021domain}
 \citep{chung2021domain} \citep{scalbert2022test} \citep{yamashita2021learning} 
 \citep{vuong2022impash}  \citep{xiong2020improve}       (Section~\ref{sec:Data augmentation}) , xnode,  l sep=10mm,
]
    ]  
    [Representation \\ learning (Section \ref{sec:Representation Learning}), root,  l sep=10mm,
       [Adversarial learning \citep{long2021domain}  \citep{wilm2021domain}  \citep{guan2020attention} \citep{chen2020cross} \citep{janizek2020adversarial} (Section~\ref{sec:Adversarial}), xnode,  l sep=10mm,
        ]
            [Feature distribution alignment \citep{meng2020unsupervised} (Section~\ref{sec:Feature distribution alignment}), xnode,  l sep=10mm,
    ]
      [Contrastive learning \citep{gurpinar2022contrastive} \citep{le2021combining}  (Section~\ref{sec: Contrastive leanrning}), xnode,  l sep=10mm,
] 
   [ Consistency regularization  \citep{zhang2022semi} \citep{raipuria2022stain} 
 \citep{li2020domain} \citep{reiter2023domain} \citep{viviano2019saliency} (Section~\ref{sec:Regularization}), xnode,  l sep=10mm, 
]
    ]
            [Learning \\ strategies (Section~\ref{sec:Learning strategies}), root,  l sep=10mm,
  [Ensemble learning   \citep{wang2023domain} \citep{philipp2022dynamic}   \citep{shen2022cd2} \citep{andreux2020siloed} (Section~\ref{sec:Ensemble learning}), xnode,  l sep=10mm, 
   ]
     [Test-time augmentation \citep{scalbert2022test} \citep{bissoto2022artifact} (Section \ref{sec:Test-time}), xnode,  l sep=10mm, 
   ]
     [Incremental \citep{seenivasan2022biomimetic} \citep{seenivasan2023task} (Section~\ref{sec:Incremental learning}), xnode,  l sep=10mm,    
]
 [Self-supervised learning \citep{li2021domain} \citep{vuong2022impash} \citep{seenivasan2023task} \citep{lee2021suprdad}   (Section~\ref{sec: Self-supervised learning}), xnode,  l sep=10mm, ]
  [Meta-learning \citep{li2022domain} 
 \citep{bayasi2022boosternet} \citep{sikaroudi2022hospital}  (Section \ref{sec:Meta-learning}), xnode,  l sep=10mm,]
      [Gradient operation  \citep{atwany2022drgen} (Section~\ref{sec:Gradient operation}), xnode,  l sep=10mm,   ]
     [Distributionally Robust optimization \citep{bissoto2022artifact} (Section~\ref{sec: Distributionally Robust optimization}), xnode,  l sep=10mm, ]
      [Multi-task learning \citep{lin2022camera} \citep{wang2022embracing} \citep{razavi2021cascade} (Section~\ref{sec:Multi-task learning}), xnode,  l sep=10mm,   
    ]    
   ]
        ] 
\end{forest}}
    \caption{Literature survey tree for covariate shift.}
    \label{fig:Figure_3}
\end{figure*}
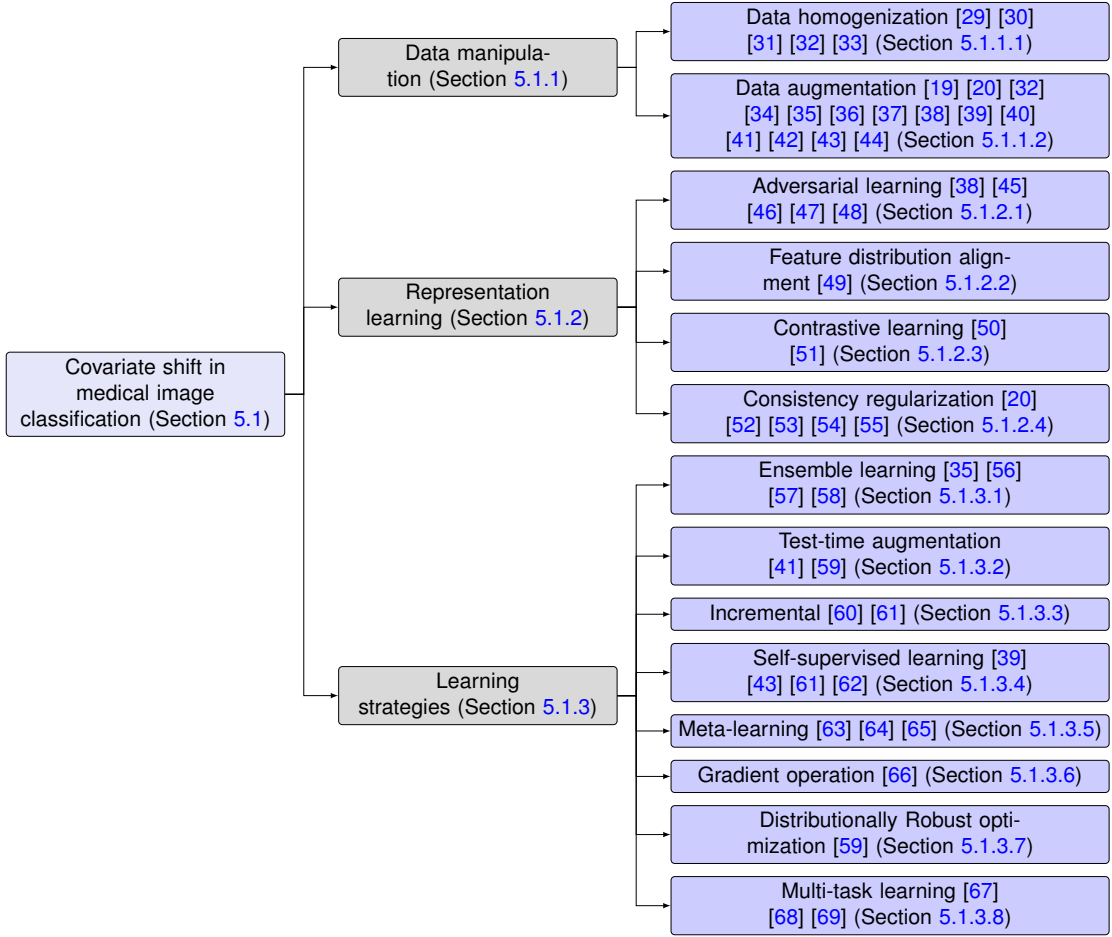

\subsection{Covariate shift in medical image classification}
\label{sec:covariate_shift}
Data heterogeneity is a key challenge for DL model generalizability. 
Covariate shift, in particular, is considered one of the most prominent shift in medicine. It is difficult to avoid this type of shift in medical imaging. It is mainly caused by the use of different type of acquisition systems and protocols, which may present notable differences among domains (i.e., changes in intensity values and contrast). Another factor to covariate shift is the differences in the characteristics of the lesions or diseases (shape, size, malignancy and location) and biological variations between patients (age, sex).
Solutions for tackling the covariate shift can be categorized into: data manipulation (Section~\ref{sec:Data manipulation}), representation learning (Section~\ref{sec:Representation Learning}) and learning methods (Section~\ref{sec:Learning methods}).

\subsubsection{Data manipulation}
\label{sec:Data manipulation}
Data manipulation methods focus on data-driven approaches to achieve robust model to domain shift, hence improve the generality of DL models. These methods can be categorized into \textit{data homogenization} and  \textit{data augmentation}. Data homogenization attempts to normalize the data and reduce the variance which exists between source domains. On the contrary, data augmentation applies augmentation techniques (severe augmentations) to expand the style variance and incorporate more diversity.

\subsubsubsection{Data homogenization}
\label{sec:Data Homogenization}
Data homogenization aims to pre-process images in a way to eliminate specific signals of each domain.

\citet{kurian2022domain} and \citet{yin2022afa} proposed to use a pre-processing anto-encoder to reduce the domain shift problem. The main idea is to produce a uniform domain appearance of input images prior to applying a classification network. The auto-encoder is trained to reconstruct the input images using a Mean Square Error (MSE) loss. To further erase domain specific signals, adversarial learning is incorporated using a domain discriminator, which is a network tasked to detect the domain label (i.e., the scanner technology used to acquire images).  On the other hand, the autoencoder is trained to maximize the domain label prediction loss and minimize the reconstruction loss simultaneously. Unfortunately, when applied for mitotic figure detection in Whole Slide Images (WSI) \citep{kurian2022domain}, this method performed very poorly with an F1 score of 0.0030 on the test set of the MIDOG 2021 challenge.
 Instead of applying the autoencoder in the spatial domain, 
\citet{yin2022afa} proposed to apply it in the frequency domain, under the assumption that the amplitude spectrum encodes the style information whereas the phase spectrum contains the content details. The goal was to learn a frequency attention map that can align different domain images in a common frequency domain. That is, the input image was first converted to the frequency domain. The phase spectrum of the input image remains unchanged. In contrast, its amplitude spectrum is reconstructed using an autoencoder which filters out domain specific frequency information. In the context of lung nodule detection from CT images, they reported a competition performance metric \citep{niemeijer2010combining} of 0.911 on the target test set of LUNA-DG.

\citet{gunasinghe2022domain} considered three classical preprocessing methods: median filter, input standardization, and randomized multi-image histogram matching. The median ﬁlter is a non-linear digital preprocessing technique, used to remove noise from an image. Input standardization, a method inspired by \citet{quellec2020automatic}, aims to attenuate illumination variations. Histogram matching is a technique that transforms the histograms of the red, green and blue channels of an image to match those of a specific reference image. In randomized multi-image histogram matching, histogram matching is performed sequentially using multiple reference images selected from the training source domain. When tested for glaucoma detection in fundus photographs using RIMONEv2 and REFUGE, the results have shown that standardization of images led to greater performances in most scenarios with an average Area Under the Receiver Operator Characteristic Curve (AUC) of 0.85. 

Inspired by \citet{nyul2000new}, \citet{garrucho2022domain} proposed to perform intensity scale standardization, a two-step technique consisting of: 1) a training step, where a standardized histogram is learned from the training images to identify key histogram landmarks, and  2) a transformation step, in which  the images are adjusted using the parameters learned in the first step. When applied for mass detection in mammography, enhanced generalization performance were achieved, outperfoming MixStyle, Cutout, RandConv and histogram equalization.

\citet{wang2021harmonization} proposed to use normalizing-flow-based method for counterfactual inference within a Structural Causal Model (SCM), to attain harmonization of data. The idea is to explicitly model the causal relationship of known confounders such as site, gender and age, and ROI features (i.e., the imaging measurement) in a SCM which uses normalizing flows to model probability distributions. Counterfactual inference can be performed upon such a model to sample harmonized data by intervening upon these variables. For the task of age regression and Alzheimer's disease classification, this method  showed better cross-domain generalization compared to state-of-the-art algorithms such as ComBat and IRM, and to models trained on raw data. 

\subsubsubsection{Data augmentation}
\label{sec:Data augmentation}
While data-augmentation in DL is used to prevent overfitting on the training set and improve in-domain generalizability, when applied in the context of DG, it aims to improve the DL generalizability to unseen target domains. Therefore, the generated samples in DG may be visually different to those in the source domain, in contrast to typical synthesized images \citep{li2022single}.

In this context, \citet{li2022single} proposed \textit{Amplitude Spectrum Diversification} for single-DG to improve the diversity of training data. First, an input image is converted into the frequency domain using the Discrete Fourier transform. Then, diverse samples are generated by modifying the amplitude spectrum using a variety of randomization operations, i.e., randomize the amplitude and position of points in the amplitude spectrum using rescaling and pixel shuffling operations. One advantage of their proposed method is that no extra network is needed for adversarial sample generation. The authors reported an average accuracy over all out-of-domain data of 0.6285 for the MIDOG dataset and of 0.6287 on a multicenter colposcopic image dataset.

\citet{zhang2022semi} integrated a similar strategy, using \textit{domain randomization}, which was implemented using \textit{amplitude mix} or \textit{color jitter}. In \textit{amplitude mix}, an image is perturbed through linearly interpolating its amplitude spectrum with that of another image.  In \textit{color jitter},  variations are introduced in terms of hue, saturation and contrast distributions.

\citet{lucieri2022revisiting} presented \textit{Amplitude-Focused Amplitude-Phase Recombination for pair samples (AF-APR-P)}\footnote{\href{https://github.com/adriano-lucieri/shape-bias-in-dermoscopy}{https://github.com/adriano-lucieri/shape-bias-in-dermoscopy}}. It aims to enhance the model's ability to generalize by focusing on the amplitude spectrum of the images while altering the phase spectrum. This is achieved by swapping the phase spectrum among images but retaining their original amplitude spectrum. The authors showed improved performance on binary skin lesion classification tasks on the International Skin Imaging Collaboration (ISIC) dataset\footnote{\href{https://www.isic-archive.com/}{https://www.isic-archive.com/}} and the seven-point checklist criteria dataset \citep{kawahara2018seven}.

\citet{wang2023domain} extended the conventional mixup to \textit{cross-domain mixup} to create a virtual domain based on the data from source domains. The original mixup technique produces convex combinations of pairs of images and their labels: it interpolates pairs of samples from the same domain that are drawn at random. In cross-domain mixup, one combine pairs of samples from different domains, to form a virtual domain $\mathcal{S}_{mix}$ that comprises virtual images ($x_{mix}$) and labels ($y_{mix}$), as formulated in the following equations: 
\begin{equation}
x_{mix}= \lambda x_1 + (1- \lambda) x_2
\end{equation}
\begin{equation}
y_{mix}= \lambda y_1 + (1- \lambda) y_2
\end{equation}

where $(x_1, y_1)$ and $(x_2, y_2)$ denote a pair of samples from source domain $\mathcal{S}_1$ and $\mathcal{S}_2$, respectively.
$\lambda \sim Beta(\alpha, \alpha)$ for $\alpha \in (0,\infty)$ and $Beta(\alpha, \alpha)$ is a $Beta$ distribution with two equal parameters $\alpha$ and $\alpha$. $\alpha$ is set to 0.4.

Experiments performed on chest X-rays datasets for the diagnosis of thoracic diseases  showed that their proposed method outperformed Empirical Risk Minimization (ERM) and six other DG approaches.

\citet{garrucho2022domain} also studied different augmentations techniques for DG. Namely, Cutout \citep{devries2017improved}, RandConv \citep{xu2020robust} and MixStyle \citep{zhou2021domain}. 
In addition, they investigated a data homogenization approach, the intensity scale standardization approach (presented in Section \ref{sec:Data Homogenization}). They evaluated the performances of their model using one or a combination of data augmentation strategies. The experiments for mass detection in mammography showed that the combination of intensity scale standardization and cutout data augmentation led to the best results in all unseen domains.

To enhance their model's generalizability to different devices, \citet{lafarge2021rotation} incorporated a sequence of transformations such as transposition, color shift, Gamma correction, Hue rotation, spatial shift, additive Gaussian noise and cutout \citep{devries2017improved}. The evaluation of this method for mitotic figure detection on the preliminary test set of the MIDOG challenge resulted in a F1 score of 0.6828.

\citet{dexl2021mitodet} performed a single augmentation to each image as part as a very simple random augmentation approach, inspired by Trivial Augment \citep{muller2021trivialaugment}. These augmentations are uniformly selected from a pool of color, noise, and special transformations, with the intensity of augmentation randomly chosen within a predefined range. Each image is also randomly flipped, and the RGB channels are shuffled at random. Using this strategy, their model achieved an F1 score of 0.7138 on the preliminary test phase of the MIDOG challenge.

In the domain of histopathology, \citet{long2021domain} proposed to broaden the spectrum of stain color appearances in the training images. This was achieved by introducing randomness in selecting stain normalization techniques and target color styles. This method used two \textit{stain normalization} techniques: Reinhard \citep{reinhard2001color} and Vahadane \citep{vahadane2015structure}. Each technique was applied using a specific probability. To achieve robust detection performance for variety of images, they gradually expand the color style ranges to the network until there is a degradation in the detection performance. Their model achieved an F1 score of 0.7500 on the preliminary test phase of the MIDOG challenge.

\citet{li2021domain}, \citet{chung2021domain} and \citet{scalbert2022test} proposed a \textit{GAN-based} approach to expand the style variance of the training data.
\citet{li2021domain} employed CycleGAN\footnote{\url{https://github.com/lizheren/MSVCL_MICCAI2021}} to map the images of source domain to a device-style domain. The augmented images were then used in their contrastive learning strategies to develop a representation with better generalization capability to various device domains (Section~\ref{sec: Self-supervised learning}). In addition, to enhance sample diversity, they used different diversifying operations including random cropping, random rotation, horizontal flipping, and adjustment of brightness, contrast, and saturation.

\citet{chung2021domain} adopted StarGAN\citep{choi2018stargan} to translate images into arbitrary device styles (based on the mixing of device characteristics), without losing morphological information upon training. Next, a detection network was trained on the translated images for mitotic figure detection. Their model achieved an F1 score of 0.7548 on the preliminary test phase of the MIDOG challenge.

\citet{scalbert2022test} introduced \textit{Test-Time data Augmentation (TTA)} based on StarGANV2 \citep{choi2020stargan}\footnote{\url{https://gitlab.com/vitadx/articles/test-time-i2i-translation-ensembling}}, a more recent multi-domain image-to-image translation model. The idea is to project images from unseen domain into each source domain, classify the generated images and ensemble their predictions. The proposed method has shown good results when evaluated for two different histopathology tasks: 1) patch classification of lymph node section WSIs and 2) tissue type classification in colorectal histological images. This method outperformed standard/ Hematoxylin\&Eosin (H\&E) specific color augmentation/normalization and standard test-time augmentation techniques.

In their \textit{Style Transfer Augmentation for Histopathology (STRAP)} \footnote{\url{https://github.com/rikiyay/style-transfer-for-digital-pathology}} data augmentation approach, \citet{yamashita2021learning} proposed to use image-to-image translation models at the testing phase. They employed random style transfer from non-medical style source (such as natural images from the miniImageNet dataset \citep{vinyals2016matching}) by applying AdaIn style transfer \citep{huang2017arbitrary}, as in \citet{geirhos2018imagenet}. 
That is, the style of medical images (i.e., histopathology images), namely the texture, color and contrast are translated with the style of a selected non-medical image. However, the semantic content of the image, the global object shapes are unchanged. Their method was applied for 1) colorectal cancer classification into two distinct genetic sub-types based on WSI in a single-DG setting and 2) identifying the presence or absence of breast cancer metastases in image patches extracted from histopathlogic scans of lymph node sections in a multi-source DG setting. It achieved higher performances compared to stain normalization based approaches. Despite promising performances, applying AdaIn as on-the-fly data augmentation is considered to be computationally expensive. 

With the aim to learn invariant representation, resistant to domain shift, \citet{vuong2022impash}\footnote{\url{https://github.com/trinhvg/IMPash}} proposed a new augmentation strategy called \textit{PatchShuffling}. Inspired by Pretext-Invariant Representation Learning (PIRL) \citep{misra2020self}, it is used during the pre-training phase, along with another type of augmentation \textit{InfoMin} \citep{tian2020makes}. Unlike PIRL, which starts by extracting the patch feature and then rearranging these features within the initial image, PatchShuffling directly shuffles the initial image itself. Initially, PatchShuffling randomly selects a portion from the image, ensuring its size is approximately [0.6,0.1] of the original image area. This cropped image is then resized and randomly flipped. They randomly extract 9 non-overlapping patches and assemble them as 3-by-3 grid to form a new image. On the other hand, the InfoMin augmentation constructs two views of the original image: it is designed to minimize the mutual information between the original and the augmented version of an image, while preserving any task-relevant information intact. Their framework outperformed other traditional histology domain-adaptation and self-supervised learning methods in the task of colorectal cancer tissue classification.

\citet{xiong2020improve} introduced \textit{Enhanced Domain Transformation} (EDT) for improving DG on unseen images. It incorporates several image processing steps: 1) image local average subtraction, 2)  average blurring for reducing high-frequency noise and adaptive local contrast enhancement for normalizing the images, 3) PCA color jittering which modifies the training image color with the predominant color component to simulate the color characteristics of the unseen domain. The provided image might originate from the known domain, unseen domain or even non-medical images (ImageNet, etc.). This method was applied for age regression and DR classification using fundus photographs. Despite promising results, the average blurring process can mask important features, reducing the model's classification performance.

\subsubsection{Representation Learning}
\label{sec:Representation Learning}
Representation learning involves training a parameterized model to learn the mapping from the raw input data to a feature vector, with the aim of uncovering more abstract and useful concepts. This process is designed to enhance the effectiveness of various downstream tasks by capturing the essential information embedded in the data \citep{le2020contrastive}. In the context of DG, representation learning mainly focus on  the concept of domain alignment for creating robust and generalized representations to unseen data. The goal of domain alignment is to minimize the difference among source domains for learning domain-invariant representations. It assumes that domain-invariant representation to the source domain should also be robust to unseen test domain. Recently, many methods have emerged to measure the  distance between distributions and achieve domain alignment. These methods can be categorized into four main groups: adversarial learning, feature distribution alignment, contrastive learning, and consistency regularization. 

\subsubsubsection{Adversarial learning}
\label{sec:Adversarial}

\begin{figure}[htbp!]
  \centering
  \includegraphics[width=0.8\columnwidth]{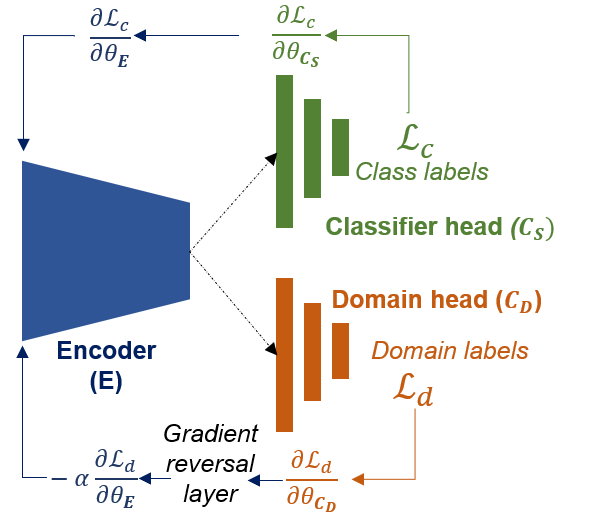}
  \caption{Adversarial learning.}
  \label{fig:Figure_5}
\end{figure}

In DG, adversarial learning is utilized to acquire source domain-invariant features that can be effectively used on new testing domains. In general, this is achieved by training an encoder ($E$) with an adversary discriminator (i.e., a domain discriminator, $C_D$) and a category classifier ($C_S$), as illustrated in Figure \ref{fig:Figure_5}. The domain discriminator is tasked to distinguish the domains of the input features by minimizing the cross-entropy loss (${\mathcal {L}_d}$). The category classifier is employed for the main task of classification (${\mathcal {L}_c}$). The end goal is to learn domain-invariant features across the source domains, that is to accurately predict disease labels without relying on any domain indicators. To this end, the feature extraction network is trained to confuse a domain discriminator and to accurately classify diseases. It is jointly trained to maximize the domain classification loss and to minimize the category classification loss:

\begin{equation}
\mathcal {L}_{total}= \mathcal{L}_c- \alpha \mathcal{L}_d
\end{equation}

where $\alpha$ is a hyperparameter to control the contribution of adversarial loss. In practice, given that the feature extraction network parameters are jointly updated by the backpropagation of the category classifier and the domain discriminator, a self-defined \textit{gradient reversal layer} is added to transmit negative gradient variations from the domain discriminator. Note that in the forward propagation, this layer acts as an identity transform.

For instance, \citet{wilm2021domain} 
 incorporated a gradient reversal layer and a domain classifier to their RetinaNet model\footnote{\url{https://github.com/DeepMicroscopy/MIDOG}} employed for object detection. To learn domain invariant feature, the network is trained using the domain classification loss for all source domains, the bounding box regression loss and the instance classification loss. For the task of mitotic figure detection, this method achieved an F1 score of 0.7183 on the MIDOG challenge’s test set.

To address the same problem, another work \citep{long2021domain} proposed an UDA technique based on adversarial training. It comprised a pretrained ResNet-50 as a backbone, three cascaded detection heads for high quality detector, and a PatchGAN discriminator. The PatchGAN discriminator is trained to distinguish features between source images and target images. At the same time, it guides the training of the network in an adversarial manner. One advantage of PatchGAN is that it can be applied to images with arbitrary sizes. In comparison to \citep{wilm2021domain}, their model achieved a better F1 score of 0.7500 for mitosis detection on the preliminary test phase of the MIDOG challenge.

\citet{guan2020attention} also presented an UDA approach based on adversarial learning, named \textit{ Attention-driven Deep Domain Adaptation}. It is composed of: 1) a feature encoding module, 2) an attention discovery module that discovers disease-related regions and 3) a domain transfer module with adversarial learning comprising two classifiers: a domain discriminator and a category clasifier. By co-training the two classifiers, the model is supposed to learn domain-invariant features for both domains (source and target domains) as well as strong classification performance for the source data, which increases the robustness of the learnt model when used for the target domain. This method showed good results for brain dementia identification and disease progression prediction when evaluated on three benchmark neuroimaging datasets ADNI-1 \citep{jack2008alzheimer}, ADNI-2 \citep{jack2008alzheimer}, and  AIBL \citep{ellis2009australian} datasets.

In their UDA adversarial framework, \textit{Cross-device and Cross-anatomy Adaptation Network}, \citet{chen2020cross} aimed to enhance anatomy classification in ultrasound video. Their main idea was to align the distribution of multi-scale deep features in adversarial training. This alignment involved training two discriminators, a local and a global discriminator, that assess whether pairs of features are a positive or negative pair from the same image based on their mutual information. The local discriminator enhances the correlation between local  convolutional features and a unified global semantic feature, while the global discriminator aligns the global semantic feature with the classifier predictions. Their approach showed promising results in ultrasound anatomy classification, with mean recognition accuracy increasing by 20.8\% and 10.0\%, compared to a method without domain adaptation and an adversarial learning-based domain adaptation method, respectively.

\citet{janizek2020adversarial}\footnote{\url{https://github.com/suinleelab/cxr_adv}} proposed an \textit{adversarial deconfounding} approach to improve pneumonia detection in chest X-rays. Their aim was to make their pneumonia detection model invariant to the view position of chest radiographs (anterior-posterior vs. posterior-anterior). This was achieved by jointly training a classifier with an adversarial network that tries to determine the view from the classifier output score. In contrast to previously mentioned methods, this approach does not require data from the target domain, instead it is based on domain knowledge about the causal relationships involved in the data to identify nuisance variables. These variables might relate differently to the outcome in the test domain then in the source domain. To overcome this issue, adversarial technique has shown promising results to train a classifier that is invariant to the nuisance variable.

\subsubsubsection{Feature distribution alignment}
\label{sec:Feature distribution alignment}
An alternative to domain adversarial learning for achieving domain-invariant representations is to match the feature distributions. This is typically achieved using information theory based technique such as  Kullback-Leibler (KL) divergence. By minimizing the KL divergence, all source domain representations are aligned with a Gaussian distribution. On the other hand, other strategies such as Minimax Entropy (MME) \citep{saito2019semi} use the principle of entropy minimization to achieve domain alignment.

In their UDA framework, \textit{MetFA}\footnote{\url{https://github.com/qingjie99/MetFA}}, \citet{meng2020unsupervised} proposed to learn a shared latent representation space between the source and  target domains using a Gaussian embedding modeled by a standard Gaussian distribution. This distribution matching is achieved through the KL divergence. Inspired by MME \citep{saito2019semi}, class representations (prototypes) are estimated in this shared latent space. These prototypes, which correspond to the weights of the last dense layer in the classifier, are initially transitioned from the source domain to the target domain by maximizing the conditional entropy of unlabeled target data. In the second step, features are clustered around these prototypes by minimizing the entropy with respect to the feature extractor. 
 Furthermore, in order to maximize the margin between different classes across domains, a cross-domain metric learning was proposed. It aims to minimize the distance between the latent features of the target data (query samples) and the latent features of the labeled source data (support samples) when they belong to the same class, while maximizing the distance when they are from different classes. 

Additionally, MetFA  aligns the class distributions between the source and target domains.
Following \citet{dou2019domain}, soft label distributions are computed for both domains using a ``softened" softmax at temperature $\tau$. The class distribution alignment loss is then assessed using the symmetrized KL divergence between these soft label distributions.
This method was evaluated for cross-device anatomical classification of fetal ultrasound view planes. It achieved an F1 score of 0.5776 and of 0.7713 on the target data coming from the GE Voluson E8 device and the Philips EPIQ V7 G device, respectively.

\subsubsubsection{Contrastive learning}
\label{sec: Contrastive leanrning}
\textit{Domain alignment} based on contrastive learning has emerged as an effective strategy. These techniques often employ contrastive learning, a machine learning paradigm which can be viewed as learning by comparing. In Contrastive Learning (CL), a representation is learned by comparing among the input samples. The comparison can be conducted between pairs of similar inputs (positive pairs) and  pairs of dissimilar inputs (negative pairs).
The method involves computing the distance between feature vectors of image pairs and deriving the loss according to this distance. An image pair is deemed sufficiently dissimilar if the computed distance exceeds a predetermined margin. Typically, most studies considered images coming from the same class as similar pairs. 
 
\citet{gurpinar2022contrastive} proposed a DA approach based on contrastive learning with cosine distance. A Siamese network is trained using a CL loss to learn embeddings such that samples from the same class are gathered closer and samples from different classes are pushed away. To further adapt it to a multi-label classification problem,  a new smoothing parameter, $\beta$ (ranging between 0 and 1), is added to the loss to make it proportional to the similarity regarding present labels. 

\begin{equation}
    \beta = \frac{|(Y_\mathcal{S}\cup Y_\mathcal{T})|- |(Y_\mathcal{S} \cap Y_\mathcal{T})|}{|(Y_\mathcal{S}\cup Y_\mathcal{T})|}
\end{equation}

where $Y_\mathcal{S}$ and $Y_\mathcal{T}$ denote the label vectors for a pair of source and target images, respectively.
The contrastive loss is then updated as: 

\begin{equation}
\mathcal{L}_{contrastive}(Y, d)= \frac{1}{2} (1- Y) d^2 (1+\beta) +(Y)\frac{1}{2} max (0,m-d)^2 
\end{equation}

where $d$ denotes the cosine distance between feature vectors extracted from the pair of images, $m$ indicates the margin and $Y$ represents the pair labels ($Y=1$ for dissimilar pairs and $Y=0$ for similar pairs). This method was applied for facial action unit detection  for children with hearing impairments. Integrating $\beta$ led to improved recognition performance with a weighted F1 score ranging between 0.76 and 0.85 on the target HIC dataset.

\citet{le2021combining} combined data augmentation approaches with domain alignment based on CL. The essence of this method is to minimize the distance between original and augmented domains. Positive pairs consist of pair of samples from the same class while negative pairs are from different classes. Augmented domains were obtained by applying techniques such as random cropping, random horizontal flip, random color jitter, and random grayscale. To enforce invariant features, the distance between original and augmented domains was minimized using a supervised contrastive learning loss in the form of normalized temperature-scaled cross-entropy loss. This method was assessed on PACS and on a medical benchmark dataset of chest X-ray, consisting of data from CheXpert, ChestX-ray14 and PadChest.
A disadvantage of this method is it assumes that the label distribution to be roughly equal across the domains, implying the need for balanced datasets. 

\subsubsubsection{Consistency regularization}
\label{sec:Regularization}
Consistency regularization methods mainly add a loss term to the learning objective  to make the model robust against variations in input data that are irrelevant to the classification task such as changes in texture, color, etc.  

\citet{zhang2022semi} proposed a semi-DG method,   which constrains the learned representation to have two characteristics: stability and orthogonality. Their regularization objective was applied to features from (labeled or unlabeled) pairs of original and domain-randomized augmented images. To enforce feature stability, the sum of the channel-wise cosine distance between the original feature and its augmented version (its domain-randomized counterpart) was computed. The orthogonality of features was assessed using the cosine similarity between different channels. This method was applied for chest X-ray diagnosis, using MIMIC \citep{johnson2019mimic} as the labeled source and NIH \citep{wang2017chestx} and CXP \citep{irvin2019chexpert} as unlabeled source domain. It was tested on the PadChest \citep{bustos2020padchest} dataset, showing promising results with a mean AUC of 0.8443 for detecting pathologies.

In the context of computational histopathology, \citet{raipuria2022stain} introduced a \textit{consistency regularization loss} to ensure their model remains highly invariant to stain color changes on unseen test data.  Their model was enforced to produce consistent predictions for both original samples and their stain modified versions using the KL divergence loss. In addition, an auxiliary task of stain regeneration was applied to enhance the model's  generalization capabilities. This involves training a decoder to regenerate the original stain color using feature representation of the stain modified images ($\hat{x}$). Thereby, a shared representation is learned for the primary task of classification and the auxiliary task of stain regeneration. This method was evaluated on two publicly available datasets TUPAC-16  and Camelyon17. It showed that stain-invariant features results in improved performance on unseen images coming from different centers.

To improve DG performances,  \citet{li2020domain} presented a rank-regularized latent feature space\footnote{\url{https://github.com/wyf0912/LDDG}}.  Based on the assumption that there are linear dependencies between the latent features of different domains, the latent feature space was regularized by modeling intra-class variation using rank constraint: the rank of the latent feature matrix was constrained to the number of classes. At the same time, the distribution of latent features was aligned to a common Gaussian distribution. This approach was evaluated for skin lesion classification using seven public skin lesion datasets. Using  ResNet18 model, it showed better cross-domain generalization performances when compared to state-of-the-art baselines. Inspired by these results, \citet{reiter2023domain} incorporated this approach in their \textit{Detection Transformer (DETR)} \footnote{\url{https://github.com/facebookresearch/detr}} model \citep{carion2020end} for 
 the purpose of DG in real-time surgical tool detection. Despite reporting improved generalization performances, the limited size of the datasets was a major limitation. 

\citet{viviano2019saliency} explored different  regularization strategies\footnote{\href{https://github.com/josephdviviano/saliency-red-herring}{https://github.com/josephdviviano/saliency-red-herring}} where the DL model is trained to ignore confounders (such as acquisition site) using attribution (saliency) priors, i.e.,  expert-drawn masks highlighting relevant regions for predictions. These methods consisted of: 
1) an \textit{activation difference approach}, which regularizes the model and penalizes the L2-normed distance between the masked and unmasked input's latent representations, 2) an \textit{adversarial approach} which employs a discriminator to identify whether latent representations come from a masked or unmasked input, and 3) two saliency penalties methods (\textit{GradMask} \citep{simpson2019gradmask} and \textit{Right for the Right Reasons (RRR)}\citep{ross2017right}) which penalize the model for producing saliency gradients outside of regions of interest.
Despite improved generalization performance in the presence of covariate shift, the results showed that the DL network still attribute features outside of the mask at test time. Indeed, the proposed methods do not guarantee to negate any confounding variables that exist within the mask.

\subsubsection{Learning strategies}
\label{sec:Learning strategies}
Different learning paradigms have been proposed to enhance generalization performances. These learning techniques include: ensemble learning, test-time augmentation, incremental learning, self-supervised learning, meta-learning, gradient-operation, distributionally robust optimization, and muli-task learning.

\subsubsubsection{Ensemble learning}
\label{sec:Ensemble learning}
In machine learning, ensemble methods are very common to boost generalization performance. The principle of ensemble learning is to derive a prediction given predictions from multiple models (i.e., an ensemble).  This is typically implemented as a simple averaging over the ensemble predictions. In DG, more generally, ensemble learning refers to combining multiple models to enhance generalization.

In the context of surgical instrument localization, \citet{philipp2022dynamic} proposed  an uncertainty-based dynamic CNN which combines two modalities (image and optic flow modality).
Their CNN dynamics were guided using pixel level uncertainty estimated separately for each modality. It comprises two ensemble network, one for each modality. The outputs for each ensemble is fused using the mean of the prediction maps. Next, pixel-wise uncertainty map is estimated using the standard deviation across the ensemble individuals. Uncertainty masks are then computed by normalizing these pixel-wise uncertainty maps.
Finally, these uncertainty masks are used to fuse the ensemble for the two modalities by weighting the predictions from each modality based on their respective uncertainties. 
This method  showed good generalization performances when evaluated on  heterogeneous surgical datasets coming from different domains including eye, laparoscopic and neurosurgeries.

\citet{wang2023domain} proposed \textit{domain-ensemble learning with cross-domain mixup}. Their model  comprised a shared backbone for all source domains and a domain-specific classifier. After training their domain specific model,  they used ensemble learning to expose the model optimization to domain distribution discrepancy. They enforced the consistency between the predictions obtained by all non-domain specific model (ensemble of predictions) and a pseudo label  generated by a domain-specific model (i.e., prediction from a domain-specific model).
This method was applied for thoracic disease classification in unseen domains and showed that it outperformed the state-of-the-art DG methods on unseen datasets.

In their Federated Learning (FL) framework, \citet{shen2022cd2}  and \citet{andreux2020siloed} addressed the non-IID data across different clients.
 \citet{shen2022cd2} presented a \textit{channel decoupling strategy} for model personalization. The network of each client ($i$) was composed of private personalized parameters $\theta_i$, and global shared parameters $\theta_0$. Their vertical decoupling strategy consisted of assigning an adaptive proportion of learnable personalized weights at each layer from the target model, moving from the top layers to the bottom layers. A uniform personalization partition rate, ranging between zero and one, was defined to determine the precise proportion of the personalized channels in each layer. To enhance the collaboration between private and shared weights, they used a \textit{cyclic distillation} scheme. For each input sample, they used the KL divergence to impose a consistency regularization between $\theta_i$ and $\theta_0$, guiding the predictions from $\theta_i$ and $\theta_0$ to align to each other. They showed that their channel decoupling framework can deliver more accurate and generalized results, outperforming the baselines when evaluated on Histo-FED dataset.

\citet{andreux2020siloed} presented \textit{SiloBN}, another model personalization method based on FL. It uses local-statistic Batch Normalization (BN) layers to discriminate between local and domain-invariant data. Only the learned BN parameters are shared across centers, whereas BN statistics (the running means and variances of each channel computed across both spatial and batch dimensions, respectively) remain local. To generalize the resulting models to unseen centers, similar to AdaBN \citep{li2016revisiting},  the BN statistics are recomputed on a data batch from the target domain while the other model parameters are kept frozen as obtained from the federated training. This approach has shown promising out-of-domain generalization performances when assessed on real-world multicentric histopathology datasets.

\subsubsubsection{Test-time augmentation}
\label{sec:Test-time}
Inspired by ensemble methods and adversarial examples, \textit{Test-Time Augmentation (TTA)} stands as a straightforward approach for estimating predictive uncertainty. This technique involves generating multiple augmented versions of each test sample by applying various data augmentation methods. These augmented images are then inputted to the model which returns an ensemble of predictions. In DG, this method can be used to project images to the source domains and then ensemble their predictions. It can also be used to select robust features for inference. 

\citet{scalbert2022test}\footnote{\url{https://gitlab.com/vitadx/articles/test-time-i2i-translation-ensembling}} integrated TTA to their DG framework based on StarGANv2. At test time, this method projects testing images to $M$ (where $M$ is the number of source domains) source domains, classify the projected images and ensemble their predictions. Given an unseen image,  $M$  style vectors are first encoded  by feeding a random latent code and its domain label to a mapping network. The StarGANV2 generator then takes the style vector and the testing image as input  to translate the image to different source domains. Experiments on different histopathology datasets showed that this method is more efficient than previous color augmentation/normalization, train-time data augmentation and DG methods. 

\citet{bissoto2022artifact}\footnote{\url{https://github.com/alceubissoto/artifact-generalization-skin}} proposed \textit{test-time debiasing}  where feature selection is performed during inference. The idea is to force the network to use the correct correlations learned to make the prediction. To reduce spurious features in testing images, \textit{NoiseCrop} was applied. It removes the background information, replaces it with a uniform noise, and resizes the lesion to occupy the whole image. For the task of skin lesion detection, this approach outperformed the baseline ERM approach and other DG methods such as RSC and GroupDRO with an AUC of 0.74 on a strong biased test set. Despite promising performances, \textit{test-time debiasing} requires domain knowledge of the task.

\subsubsubsection{Incremental learning}
\label{sec:Incremental learning}

Incremental learning also known as lifelong learning or continual learning, is a machine learning process where data arrives in sequence, or in a number of steps instead of having access to all the training data as in classical scenarios.
With the continued emergence of novel medical devices and procedure protocols,   incremental learning has gained interest in DG. It allows the model to  learn new domain shifts, without the need to retrain the model from scratch.

\citet{seenivasan2022biomimetic} proposed \textit{incremental DG}\footnote{\url{https://github.com/lalithjets/Domain-Generalization-for-Surgical-Scene-Graph}} on scene graphs to predict instrument-tissue interaction during robot-assisted surgery. They trained a feature extraction network and a graph network on a nephrectomy surgery dataset to classify 9 classes. The feature extraction network was then extended to the target domain (a transoral robotic surgery dataset) to classify 11 classes using an incremental learning technique, as described in \citep{castro2018end}. 
In addition, the authors proposed to use knowledge distillation, where the teacher network is a network trained on the source domain and the student network is a copy of the teacher network, trained on the whole target domain dataset and on a sample of the source domain dataset. To further enable the student network to retain the knowledge from the source domain while generalizing to the target domain, it was regularized using a \textit{knowledge distillation loss} between the teacher and student network logits. Despite promising performances, this method showed limited performances on the target domain.

In line with the previously mentioned paper, the authors \citep{seenivasan2023task} designed a multi-task learning model\footnote{\url{https://github.com/lalithjets/Domain-adaptation-in-MTL}} to perform tool-tissue interaction detection and scene caption. The model consists of: 1) a shared feature extractor 2) a mesh-transformer branch for scene captioning and 3) a graph attention branch for tool-tissue interaction detection.  To deal with domain shift, the authors proposed a \textit{class incremental contrastive learning} approach for surgical scene understanding. In addition, they developed Laplacian of Gaussian (LOG) based curriculum by smoothing across all three modules to enhance model learning. This approach used LOG kernels instead of Gaussian kernel to control the features entering the model at the initial epochs and highlight the instrument contours, thus allowing the model to learn gradually.

\subsubsubsection{Self-supervised learning}
\label{sec: Self-supervised learning}
Self-supervised Learning (SSL) aims to construct robust image representations via pretext tasks that do not require semantic annotations, leveraging the structure within the data itself.  Within the context of DG, SSL based on Contrastive Learning (CL) has recently emerged as a pre-training strategy to  produce generalized, pre-text invariant representations. The pre-trained model can then be adopted for various downstream tasks.

For instance, to achieve invariant representations in their SSL framework\footnote{\url{https://github.com/lizheren/MSVCL_MICCAI2021}}, \citet{li2021domain} employed two types of CL: 1) a multi-style CL to generalize to multiple device style and 2) a multi-view CL to learn representations that are robust to the  CC and MLO views in mammography. For multi-style CL, a CycleGAN was utilized to create multiple device-style images from a single source image. Positive pairs were formed by randomly selecting two images derived from the same source image. For multi-view CL, the CC and MLO views of the same breast were considered positive pairs. Following the pre-training stage, the backbone was used for the main task of lesion detection. The proposed method was assessed with mammograms from four vendors and one unseen public dataset (INbreast). It has shown significant improvement for lesion detection on both seen and unseen domains.

For the application of surgical scene understanding, \citet{seenivasan2023task} proposed to use a hybrid approach combining self-supervised learning scheme and supervised learning. 
Inspired by \citet{xu2021class}, they integrated supervised contrastive loss, also known as SupCon loss~\citep{khosla2020supervised}.
Similar to self-supervised contrastive learning, this technique applies extensive augmentation to the input and maximizes the mutual information for different views. However,  it also leverages the label information: it minimizes the distance between the same label inputs across domains and pushes apart the samples with different labels in the feature embedding space.

\citet{vuong2022impash} employed \textit{Momentum Contrast (MoCo)} \citep{he2020momentum}, which uses CL as dictionary look-up: an encoded “query” (image) should be similar to its matching key and dissimilar to others. In MoCo, a dynamic dictionary is implemented with a queue and a moving-averaged encoder (momentum encoder). The dictionary keys are defined on-the-fly by a set of data samples and are encoded by a momentum encoder. \citet{vuong2022impash} used this concept and designed two dedicated momentum branches for both InfoMin and PatchShuffling augmentations. Each branch encodes and stores a dictionary of image representations for the corresponding augmentation. The network was optimized using an extended version of InfoNCE loss \citep{oord2018representation}. When evaluated on unseen dataset for  colorectal cancer tissue classification, this approach outperformed other SSL methods such as MoCoV2 which uses a single momentum branch.
 
For boosting representation learning and improving the recognition of low-prevalent diseases, \citet{lee2021suprdad} integrated a SSL framework based on a rotation pretext task. The images in the dataset  were augmented by creating four rotated copies from $x$ by $\ 0^{\circ}, 90^{\circ},180^{\circ},$ and  $270^{\circ}$ degrees. An auxiliary head was then tasked to predict the rotation. This approach demonstrated superior performance in detecting ocular diseases in color fundus photographs, achieving a mean AUC of 96.6\% compared to 94.8\% obtained with a purely supervised learning baseline.

\subsubsubsection{Meta-learning}
\label{sec:Meta-learning}
Meta-learning, also known as learning to learn, is a paradigm aiming to learn from episodes derived from related tasks to enhance the efficacy of future learning. It has been applied to DG, by adopting an episodic training paradigm, where at each iteration a meta-task is generated with the source domains splitted into meta-train and meta-test domains to simulate domain shift. 
 
To address the problem of DG with limited data, \citet{li2022domain} proposed a mixed task sampling strategy where the meta-test domains were generated by interpolating among all the source domains. In their meta-objective, a regularization was incorporated to enforce the alignment of embeddings across training domains from both sample-wise and prototype-wise perspectives. Sample-wise alignment reduces intra-class distances while increasing inter-class separations using CL and cosine distance based loss. Domain-general prototypes were the weight vectors of the classifier, and  domain-specific prototypes were the centroid of the embedding for same-class samples for each domain. A prototype-wise alignment based on KL divergence was proposed to enforce the prediction scores across different prototypes to be consistent with each other. This approach outperformed the ERM baseline approach raising the accuracy from 88.43\% to 91.77\% on average for epithelium-stroma classification using  histopathological images.

\citet{bayasi2022boosternet} proposed \textit{BoosterNet}, an auxiliary network that can be added to any arbitrary core network to enhance its generalizability without the need to change its training procedure or its architecture. Their approach combats shortcut learning using the concept of feature culpabiblity. It uses episodic learning to learn from the most culpable features in the core network (i.e.,  features which are linked with erroneous predictions) and from the most predictive characteristics of the data (discriminant features). BoosterNet was validated for detecting skin lesions, where it showed improved generalization performance compared to other benchmark DG approaches including data augmentation based DG, adversarial training, and feature alignment.

 Inspired by \textit{Model Agnostic Learning of Semantic Features (MASF)}, \citet{sikaroudi2022hospital}  proposed to learn a latent space representation suitable for generalization to an unseen test domain. Their meta-objective was a weighted sum of an alignment loss and a triplet loss.
The alignment loss was the KL divergence between the soft confusion matrix of different domains. The metric loss was the average triplet loss for a batch of triplets, which is formed from an anchor, positive and negative instances from all the source domain dataset.  The triplet loss compares a reference input (an anchor) to a matching input (positive instance belonging to the same class as the anchor) and a non-matching input (negative instance belonging to a different class than the anchor). The distance from the anchor to the positive instances is minimized, while the distance form the anchor to the negative instances is maximized.  For the task of renal cell carcinoma subtypes classification in WSI, this method outperformed the baseline, which involved training using only cross-entropy loss on three hold-out trial sites.

\subsubsubsection{Gradient operation}
\label{sec:Gradient operation}
Some DG strategies focus on operating on gradients to develop robust models with generalized representations.   

In order to reduce gradient variance from different domains, \citet{atwany2022drgen} presented \textit{Stochastic Weighted Domain Invariance}\footnote{\url{https://github.com/BioMedIA-MBZUAI/DRGen}}, a method leveraging the Fishr regularization coupled with iteration-wise avergaging of weights (SWA).
It is built upon \textit{Stochastic Weight Averaging Densely} (SWAD) \citep{cha2021swad} and \textit {Fishr} \citep{rame2022fishr}  to encourage seeking a flatter minima while imposing a regularization.  SWAD seeks a flat minimum by averaging the weights by iterations (rather than by epochs). It enables averaging weights only from specific iterations where the validation loss decreases. On the other hand, \textit{Fishr} \citep{rame2022fishr} is a regularization approach that enforces domain invariance in the space of the gradients of the loss. In particular, the domain-level variances of gradients are matched across training domains. Fishr regularization enforces the domain-level gradient invariance in the classifier by aligning the gradient covariances at the domain level.
The Fishr loss is thus formulated as follows: 
\begin{equation}
\mathcal{L}_{Fishr}= \frac{1}{M} \sum_{i=1}^M ||cov_i-\hat{cov} ||^{2}_{F} 
\end{equation}
where $cov_i$ denotes the covariance matrix for each $\mathcal{S}^{i}$ domain for $i=\{1,...,M\}$ and $\hat{cov}$ is the mean covariance matrix, $\hat{cov}= \frac{1}{M} \sum_{i=1}^M cov_i $

The proposed method was evaluated for DR detection in fundus photographs using leave-one-domain-out cross-validation. On four public datasets (EyePACs, Aptos, Messidor and Messidor2), it achieved an average accuracy of 70.47\% compared to 62.32\% with the ERM approach.
\subsubsubsection{Distributionally Robust optimization}
\label{sec: Distributionally Robust optimization}

Distributionally Robust Optimization (DRO) \citep{sagawa2019distributionally} attempts to learn a model at worst-case distribution scenario. In comparison to ERM which minimizes the global average risk, DRO minimizes the maximum risk for all groups (or domains). This enforces the model to focus on high-risk groups, which usually comprise those with correlations underrepresented in the dataset. The risk in DRO is computed as follows: $R_{DRO}= max_{i \in\{1,...M\}} \mathbb{E}_{\mathcal{S}_i}[\mathcal{L}(x,y,\theta)]$ \citep{sagawa2019distributionally}.

\citet{bissoto2022artifact} proposed a DG approach based on DRO. Their pipeline involved first partitioning the data into training and test sets, with amplified correlations between artifacts and class labels (malignant vs. benign), which appear in opposite directions in the dataset splits. 
The training set was then divided into artifact-based domains. Next, the \textit{GroupDRO} \citep{sagawa2019distributionally} algorithm, which minimizes the loss of the worst-case training source environment, was then trained on these artifact-based  domains.
In the last phase, the authors employed a test-time debiasing procedure to reduce the influence of spurious features in the inference images. 
The experimental results for skin lesion detection showed that GroupDRO allows learning more robust features.    

\subsubsubsection{Multi-task learning}
\label{sec:Multi-task learning}
Multi-task learning is a learning paradigm where models are  jointly optimized  on several related tasks. In DG, the premise is that the model's generalization performance on classification task should be enhanced by learning robust representations, that are shared  among different tasks.
Therefore, multi-task learning can be viewed as a strategy for domain alignment, it makes possible learning of generic features by sharing parameters.

\citet{lin2022camera} proposed a multi-task network\footnote{\url{https://github.com/linzhlalala/CVD-risk-based-on-retinal-fundus-images}} for cardiovascular disease risk (CVD) estimation using fundus photographs. To learn invariant representations, a Siamese network was pre-trained using the left and right fundus photographs for each patient as positive sample pairs. 
This network was then jointly trained on WHO-CVD score and on seven clinical variables explicitly correlated with WHO-CVD such as age, systolic blood pressure and gender. They also integrated a feature-level knowledge distillation. Given two input images (left and right fundus photographs) from a single patient, the feature with the smallest supervised learning loss is considered as the teacher whereas the other as the student. For the teacher-level features, they performed stop-gradient operation when updating the feature extractor.
The results showed that the pre-training strategy reduced the feature-space discrepancy between the UK biobank dataset collected using the Topcon 3D OCT-1000 MKII and the other cameras (Mediwork portable camera).

 \citet{wang2022embracing} used the same approach for UDA by integrating auxiliary task (predicting age, gender and race) to their framework.  Their method consisted of four stages: 1) pre-training a classifier with a feature extractor, an auxiliary task network and a primary task network using source data, 2)  fine-tuning the feature extractor and the auxiliary task network for the auxiliary tasks using  target data while constraining the feature extractor not to change significantly, 3) fine-tuning the primary classifier on the source data to correctly classify the primary task (i.e., classification) based on the modified features, 4) performing inference using updated weights on the target data. The authors showed improvement of performances when tested on 3D brain MRI dataset for classifying Alzheimer's disease and schizophrenia.

For MIDOG challenge, \citet{razavi2021cascade} proposed a \textit{multi-stage mitosis detection method} based on \textit{a Cascade R-CNN}. 
The Cascade-RCNN comprises a sequential detectors with increasing intersection over union to reduce false positives. It consists of two-stage: 1) a region proposal network that detects candidate region and 2) a region proposal network  and a classification network that performs classification on the candidate regions. This method achieved a F1 score of 0.7492 on the MIDOG testing set.
This ends our presentation of DG solutions to address covariate shift.

\subsection{Concept shift and label noise in medical image classification}
\label{sec:concept_shift}
The quality of annotations has a crucial role on the model generalizability. Nevertheless, the annotation process can be subjective, and the issue of label noise is sometimes unavoidable. In addition, the quality of annotations can differ among various annotators. To improve prediction performances, DL methods have been proposed to address the problem of noisy labels and concept shift. These methods can be categorized into two main classes: \textbf{data-centric} methods and \textbf{model-centric} methods. Data-centric methods focus on \textit{data adjustment and transformation} (Section~\ref{sec:Data adjustment and transformation}). These methods focus on identifying noisy samples and correcting their labels. Model-centric methods include \textit{learning methods} (Section~\ref{sec:Learning methods}) and \textit{collaborative methods} (Section~\ref{Sec:Collaborative and iterative methods}). Learning methods propose an optimization framework based on loss functions (i.e., regularization), architecture (e.g., graphs) and learning strategies (active learning, zero-shot learning, meta-learning, etc.). On the other hand, collaborative methods exploit the cooperation between models to boost DL performances.

\subsubsection{Data adjustment and transformation}
\label{sec:Data adjustment and transformation}
Data adjustment and transformation based methods are proposed to mitigate the problem of inconsistency of medical data annotation. 
These methods focus on adjusting the labels using techniques such as sampling, label correction, data programming, or label propagation. 

\subsubsubsection{Sampling}
\label{sec:Sampling}
Sampling-based methods aim to identify samples with inaccurate labels and then proceed to either correct these labels or remove the samples entirely.

For instance, \citet{son2021leveraging} proposed to detect mislabeled samples based on the classifier's confidence (i.e., the softmax outputs) and to mask the loss function computed over such samples\footnote{\url{https://bitbucket.org/woalsdnd/codes_and_data}}.  Noisy labels were simulated in the training data by randomly flipping labels with probabilities ranging between 0 and 0.8.  To detect mislabeled samples, a filtration network  was trained on top of the classification network to minimize the logistic regression loss over validation data (i.e.,  clean data). It examines the training set with positive labels and assigns high values on positive images with clean labels and low values to suspected negative images.
For the task of referable DR detection, this method outperformed state-of-the art methods such as S-model and Bootstrap at noise ratios of 0.2 and 0.4 on the Kaggle 2015 dataset. In addition, integrating the classifier's confidence in an active learning scheme showed good results ranking first on the PALM challenge with an AUC of 0.9993 for pathological myopia classification. 

\citet{xue2022image} used confident learning \citep{northcutt2021confident} to identify noisy samples in the training data. It employs the predicted probability outputs (self-confidence) and the noisy labels for estimating label uncertainty, i.e., the joint distribution between the noisy and true labels. The class imbalance and heterogeneity in predicted probability distributions across classes are addressed by using a per-class threshold (expected self-confidence for each class) when calculating the confident joint. To further enhance the precision, \citet{xue2022image} proposed an ensemble strategy consisting of training three different classification networks and selecting the candidates that were jointly identified as noisy using confident learning. For  automated visual evaluation for precancer screening, they achieved  a kappa score of 0.687 with the cleaned development set, compared to 0.682 with the original noisy development set.

 \citet{aljuhani2022uncertainty} presented \textit{Uncertainty-Aware Sampling Framework (UASF)}\footnote{\url{https://github.com/machiraju-lab/UA-CNN}} to tackle the problem of weak labels in digital pathology, where  WSI-level diagnoses lack precise annotations indicating specific regions within WSI responsible for the diagnosed label. This method employs an informative sampling algorithm to select the most relevant tiles by estimating uncertainties using variational Monte Carlo inference, with the predictive entropy as a measure. The relevant tiles are identified by their high prediction probability and low uncertainty. Once the disease-representative tiles were effectively identified, the prediction performance was enhanced by training the model on the refined training dataset.  For the leiomyosarcoma  histological subtype grading task, this approach achieved 83\% accuracy.

\citet{bai2021cnngeno} adopted a \textit{convolutional bootstrapping strategy}\footnote{\url{https://github.com/BRF123/Cnngeno}}  to handle noisy labels in the data. First, a set of highly reliable seeds (i.e., a subset of samples) were manually selected as training set and the model was trained until convergence. The model was then used to classify the remaining samples in the dataset. The process involves selecting samples with higher classification confidence into a seed set, based on a classification confidence level set at 0.8. The expansion of the seed set is repeated until no new seeds are added, and the final trained model is then used for classification. For the calling of
structural variation genotype, the proposed method performed better than the current state-of-the-art methods on complex real data with high and low coverage.

\citet{xu2022meta, hu2022multi} proposed a \textit{sample re-weighting} algorithm that assigns weights to training samples, with higher weights assigned to clean samples and lower weights to noisy samples. In \citep{xu2022meta}, these weights are determined to minimize the loss on a clean unbiased validation dataset.  The authors showed good performances when evaluating their method on calcium imaging data of anterior lateral motor cortex, with an F1 score and balanced accuracy greater than 0.85, despite noise levels varying between 9\% to 52\%.
 In contrast to this method, \citet{hu2022multi}\footnote{\url{https://github.com/TwistedW/MIAV}}  used the concept of sample interaction in small groups as in \citet{peng2020suppressing} which does not require a clean validation set. For the automated classification of retinal arteries and veins, they achieved an accuracy of 97.47\%, 96.91\%, 97.79\%, and 98.18\% on AV-DRIVE, HRF, LES-AV and a private dataset, respectively. 

\subsubsubsection{Label correction}
\label{sec: Re-labeling}
Label correction methods focus on adjusting (re-labeling) the labels of suspected noisy samples.

To leverage incomplete observations, 
\citet{hermoza2022censor}\footnote{\url{https://github.com/renato145/CASurv}}
used the concept of pseudo labels, where the output of the network is used to estimate the label. The authors argue that the quality of generated pseudo-labels depends on the training procedure stage: during the first epochs, the pseudo-labels are less accurate than those at the last epochs. To address this issue, they used a cosine annealing schedule to control the generation of pseudo-labels during training. 
The evaluation of their proposed method on pathology and X-ray images from the TCGAGM and NLST datasets showed good prediction survival accuracy on both datasets.

Inspired by epistemic uncertainty \citep{kendall2017uncertainties}, \citet{bai2021novel} proposed Pseudo-Labeling based on Adaptive Threshold (PLAT) to reduce the generation of noisy labels in a semi-supervised approach. Unlabeled images are inputted to the model $k$ times using Monte Carlo Dropout, resulting in  $k$ predictions. Uncertainty of pixels is estimated by computing  the variance of these predictions, and then normalized by dividing by the largest variance among all predictions. The normalized result is used as an adaptive threshold.
Compared to model trained only on labeled images, this method showed a gain of 9\%-13\% in terms of F1 score. 

\citet{qiu2023hierarchical} proposed a self-training strategy consisting of noisy label cleaning optimization. Initially, the model is pretrained with the noisy labels using a large fixed learning rate, under the assumption that the network can avoid overfitting to the label noise. Then, noise-free labels (soft labels) are computed using the softmax output of the pretrained model. During each iteration, the soft labels are fixed to update the model parameters; then the model parameters are fixed, and the soft labels are updated for the next iteration.
This method achieved good performances for pathology image classification.

\citet{he2022reducing} proposed a  re-labeling module in their \textit{Self-Adaptation Network (SAN)}. For each image, they computed the softmax probabilities. Then, the maximum value of the predicted probability is compared with the probability value of the provided label in the dataset. If the predicted probability is greater than the probability of the given label by a threshold value, the sample is assigned a new pseudo-label, which is based on the model's prediction. Extensive experience on AVEC2013 and AVEC2014 demonstrated the efficiency of their proposed method for automatic depression detection. 

\citet{zhu2021hard} employed a \textit{hard sample aware self-training}\footnote{\url{https://github.com/bupt-ai-cz/HSA-NRL/}} strategy to correct and update labels. They used the mean prediction value of the sample training history of their classification model to separate the data into easy, hard and noisy samples. Their classification network was first trained on noisy data. Easy samples were identified as those with have higher mean prediction probabilities. After selecting the easy samples, noisy samples were simulated by injecting noise to the easy samples and the classification model was retrained on the simulated noisy data. Based on the mean prediction probability value, clean samples of the noisy data were identified and the rest was utilized for training a multi-layer perceptron classifier, which was designed to distinguish between hard and noisy samples using the training history of the initial classifier as input. The  classification model was then retrained on the easy and hard samples. After this step, the labels of hard and noisy samples are corrected using the pseudo-labels produced by the classification model, which correspond to the class with the highest probability model output. These steps are repeated to further purify the dataset. Finally, in the post-processing step, the noisy data with unchanged labels  and the hard samples with changed labels were dropped out. 

\citet{zhu2023robust}  included  a consistency-based noisy label correction module in their framework to detect noisy labels and correct them. It is a two-stage algorithm: 1) two networks are used to select clean samples according to their loss ranking, samples with the smallest loss are considered to be clean samples, 2) among the remaining suspected noisy data, samples that have consistent predictions on both networks are corrected. The new label (pseudo-label) is assigned as the class that both networks most strongly agree upon, under the condition that their prediction confidence surpasses the predetermined threshold.

\subsubsubsection{Data programming}
\label{sec:Data programming}
Creating large labeled datasets is expensive and challenging in some applications. To address this issue, \citet{ratner2016data} introduced data programming, a paradigm for the programmatic creation of training sets in weak supervision. It uses a generative modeling step to create weak training labels by combining unlabeled data with heuristics provided by domain experts that may overlap, conflict, and be arbitrarily correlated.

Inspired by this concept, \citet{dunnmon2020cross} proposed a framework for applying data programming to address the problem of cross-modal weak supervision in medicine, wherein weak labels derived from an auxiliary modality (text) are used to train models over a different target modality (images).
In their proposed cross-modal data programming, users provide two inputs: 1) unlabeled cross-modal data points (i.e., an imaging study and the corresponding text report), 2) a set of Labeling Functions (LFs),  which are user-defined functions
(pattern-matching rules, existing classifiers) that take in an auxiliary modality data as input (e.g., text reports) and either output a label or abstain. In the phase of offline model training, these LFs are employed on unstructured clinical reports to be combined and produce probabilistic (confidence-weighted) training labels for training a classifier on the target modality (radiograph).  Then, a discriminative text model, for instance, a Long Short-Term Memory (LSTM) network, is trained to align the raw text with the output of the generative model. 
They employed a simple heuristic optimization to determine if it is more efficient to train the final model of the target modality directly with the probabilistic labels from the generative model or if the model's performance could be enhanced by using the probabilistic labels from the trained LSTM.
During test time, the final model only takes input from the target modality and provides predictions.

This framework presents a powerful approach for reducing the reliance on hand-labeled datasets. It has shown promising results when applied to different applications spanning radiography, CT, and EEG. However, it also brings challenges related to dependence on auxiliary data, the quality of labeling functions, and potential biases.

\subsubsubsection{Label propagation}
\label{Sec:Label propagation}
Label propagation allows to take advantage of the few labeled samples to automatically annotate unlabeled samples. Given a dataset with a large number of unlabeled samples and a small number of labeled samples, this approach is based on estimating a probabilistic transition matrix that depends on the neighborhood size and a quality threshold.

\citet{vindas2022semi} proposed to estimate this transition matrix trough K-Nearest Neighbor (KNN) and local quality measures. Their approach involves four steps. First, features are extracted using an auto-encoder in an unsupervised manner. Second, t-distributed Stochastic Neighbor Embedding (t-SNE) algorithm was used to project the features into a 2D space. In this step, the optimal projection was selected based on the silhouette score. Note that only labeled samples are used for this computation. Third, the labels of high-quality labeled samples were propagated to high-quality unlabeled samples using KNN strategy and local quality metrics. This allows to increase the size of the training set. Fourth, for classification purposes, to compensate for the noise introduced by the automatic label propagation, a robust loss function \textit{a generalized cross-entropy loss \citep{zhang2018generalized}}, was introduced as follows:
\begin{equation}
\mathcal{L}(f(x), y_i)= \frac{1- f_i(x)^v}{v}
\end{equation}
where $y_i$ and $f_i(x)$ are the i-th components of the true label $y$ and the predicted label $f(x)$, $v$ is a hyperparameter which allows control of the noise tolerance and the convergence speed; when $v \rightarrow 1$ we get the mean absolute error loss function whereas when  $v \rightarrow 0$ we get the cross-entropy loss function. In their framework, $v$ was set to 0.7.
This framework was evaluated on three tasks: emboli classification, organ classification and digit classification.

\citet{ying2023covid} proposed a \textit{noisy label recovery algorithm based on Subset Label Iterative Propagation and Replacement (SLIPR)} for dealing with noisy labels in COVID chest X-ray images classification. This algorithm aims to recover label and train the CNN on the label-recovered training set. The first stage of their framework is a feature extraction and classification phase where they utilize a low-rank representation and a neighborhood graph regularization to extract both global and local features of the samples and KNN for classification purposes. The second stage consists of multi-level propagation and replacement of labels. In this stage, the concept of label propagation is used to select and replace the labels of the samples. In addition,  a selection strategy for high confidence samples was introduced. Inspired by majority voting, it selects high confidence samples as the training set based on the sample optional labels: a sample is considered to be high confidence sample if the majority result suggests that it should belong to the same type of label. 

\subsubsection{Learning methods}
\label{sec:Learning methods}
Learning-based methods are among the most used strategies to overcome the problem of noisy datasets and improve the generalizability of DL networks.
They use an optimization framework to enhance the robustness of DL networks.  
These methods include distance metric learning, active learning, zero-shot learning, gradient, robust loss function, graph and meta-model.

\subsubsubsection{Distance metric learning}
\label{sec:Distance metric learning}
Distance Metric Learning (DML) aims to learn a discriminative embedding in which similar samples are closer together, and dissimilar samples are separated \citep{dutta2021semi}. DML emerged from the concept of contrastive loss, which turns this principle into a learning objective \citep{balestriero2023cookbook}. The contrastive loss in DML captures the relationships among samples: it trains a Siamese Network, which consists of two identical subnetworks whose architecture, configurations, and weights are the same, to predict whether two inputs are from the same class. This is achieved by putting their embedding close to each other (for the same class) and far apart  (for different classes) \citep{balestriero2023cookbook,dutta2021semi}.

\citet{zhang2022re} used similar finding retrieval based on DML to improve DL models' generalizability. They employed an extra ``clean" dataset with pathological-proven labels (the SCH-LND \citep{zhang2020learning} dataset) to re-label a noisy dataset, the Lung Image Database Consortium and Image Database Resource Initiative (LIDC-IDRI) \citep{armato2011lung}. Two re-labeling methods were explored: 1) a  nodule classifier pre-trained on LIDC data and fine-tuned on SCH-LND data for malignancy labeling, and 2) a metric-based network (Siamese Network) to rank top nodule labels by computing correlations between nodule pairs.
 The Siamese Network was trained on randomly selected pairs of images from the clean dataset using the contrastive loss. During the re-labeling phase, each sample from the SCH-LND dataset was paired with an ``under-labeled" sample from LIDC, and these under-labeled samples were sorted based on their similarity scores. The new label for each sample in the LIDC dataset was obtained by averaging the labels of the top 20\% of its partner samples with the highest similarity scores. 
According to this study, relabeling through metric learning outperformed the general supervised model, suggesting that the input pairs produced by random sampling provide a data augmentation effect to learning with limited data.

 \citet{van2022differential} employed a DML-based approach within the differential learning approach to address observer-variability in training labels. This method involved training the model on an auxiliary comparison task -- determining whether a clinical parameter differed significantly  between two patients--  considered easier task and less subjective. A Siamese Network was employed to compare the estimated clinical parameter based on the generated representations. The approach showed good results in assessing left ventricle measurements in echocardiography cine series. Differential learning was integrated as an auxiliary task by computing whether there is a significant difference in Ejection Fraction (EF) between two patients (normal vs severe EF). It showed enhanced performances when evaluated on two datasets: a large cart-based dataset consisting of 28,577 echo cines obtained from 23,755 patients and 51 echo cines acquired from 23 heart-failure patients using a Point-Of-Care Ultrasound.

While previous methods addressed noisy labels due to subjective interpretations, \citet{seibold2022breaking} tackled inconsistent labels generated from unstructured medical reports via text classifiers. They proposed a contrastive language-image pre-training on report-level approach using a global-local dual-encoder architecture to learn concepts directly from unstructured medical reports and perform free-form classification. Unlike previously mentioned methods, they combined DML with self-supervision by integrating SimSiam \citep{chen2021exploring}. Two augmented versions of the same input image were created and then processed through a backbone network, an encoder-head, and a prediction-head to enforce similarity between the two views. Furthermore, they use the augmented images from the pre-training objective to mirror their text-image objectives to the augmented samples. This approach  matched the performance of direct label supervision on large-scale chest X-ray datasets (MIMIC-CXR \citep{johnson2019mimic}, CheXpert \citep{irvin2019chexpert}, and ChestX-Ray14 \citep{wang2017chestx}) for disease classification.

\citet{kurian2022improved} employed a DML method combined with Self-Supervised Learning (SSL), based on a contrastive learning framework and feature aggregating memory banks. The method comprises three phases. The first phase, the warm-up phase, uses both cross-entropy loss and contrastive loss.  A maximum loss miner function, using the contrastive loss, identifies ‘hard-pairs’ and noisy labels based on cosine similarity. The subset with maximum contrastive loss represents the hard pairs, while all the other feature vectors are updated class-wise into a fixed-size memory bank. The second phase, the weight calculation phase, assigns weights to each training sample  based on their cosine similarity to features in the memory bank, which represent clean samples. K-medoids for the features in the memory bank were also found to compute the cosine similarity for the samples with the medoids. The third phase, the final classification training phase,  involves training the model with a weighted cross-entropy loss, applying the computed similarity scores as weights.

\subsubsubsection{Active learning}
\label{sec:Active learning}
Active learning can be combined with noisy labels to enhance DL performances.

\citet{son2021leveraging} proposed a strategy\footnote{\url{https://bitbucket.org/woalsdnd/codes_and_data}} to improve the detection of a rare disease by assigning ``normal" pseudo-labels to a large number of publicly available unlabeled images. This set was combined with a small set of labeled images with the targeted rare disease for initial training. Noise was introduced in the pseudo-labels since some of the pseudo-labeled  ``normal" images likely contained the disease. Initially, their model was trained on the pseudo-labeled dataset. It was then used to identify rare disease images with high confidence predictions (greater than 0.5), effectively filtering out noise by focusing on cases where the model has high confidence.  This process significantly reduces the number of images to be manually reviewed for the rare disease. The active learning process allows to screen for the positive selected cases and correct the initial noise introduced by pseudo-labeling. The refined dataset was used for final training, achieving an AUC of 0.9993 on the PALM competition, ranking first on the off-site validation set.

\subsubsubsection{Zero-shot learning}
\label{Zero-shot learning}
Zero-shot learning (ZSL) is a technique enabling machine learning algorithms to recognize objects belonging to new, unseen classes, with the help of semantic descriptions.  A pragmatic version of ZSL is the Generalized Zero-Shot Learning (GZSL), where the test data may originate from either seen or unseen classes.

\citet{paul2021generalized} proposed a GZSL for the diagnosis of chest radiographs using a Multi-View Semantic Embedding (MVSE) network, integrating semantic spaces from X-ray reports, radiology reports, and visual traits used by radiologists. They employed a two-branch autoencoder for semantic embeddings into X-ray and CT semantic spaces. Each branch was supplemented with a  guiding network leveraging the trait-based semantic space. To improve performance for unseen classes, a self-training strategy is employed. This involves creating a self-training set of unlabeled X-ray images from seen and unseen classes. The self-training is executed in two steps: initial inference and model fine-tuning. Initially, class probabilities for unlabeled images from the self-training set are computed using the trained MVSE network. Images are then selected for both seen and unseen classes based on the highest confidence scores. Subsequently, the model is fine-tuned with this selectively chosen data for each class. This refined model is then deployed for generalized zero-shot diagnosis of chest X-rays. During testing, for a given X-ray image, the model computes distances in both the X-ray and CT semantic spaces from the respective class signatures, dynamically balancing the importance  of each branch to determine the final class probability.
This model demonstrated robust generalization capabilities on the NIH Chest X-ray dataset (NIH), a hand-labeled subset of NIH dataset (NIH-900), Open-i dataset, PubMed Central dataset (PMC) and the CheXpert dataset.

\subsubsubsection{Gradient}
\label{sec: gradient}

The \textit{Balanced Gradient Contribution} (BGC) strategy is a training approach designed to manage the significant statistical differences between domains \citep{ros2016training}. This method addresses the issue of large variance in gradients due to the distinct nature of data from each domain. In the context of DG, the BGC method could be employed to balance the learning from different domains by adjusting the contribution of gradients from each domain during the training process.

\citet{elbatel2022seamless} integrated BGC into their \textit{Seamless Iterative Semi-supervised correction of imperfect labels (SISSI)}\footnote{\url{https://github.com/marwankefah/SISSI}}, which trains object detection models with noisy and missing annotations. They introduced a range of image processing and deep learning methods to make iterative label correction. 
Using a domain adaptation strategy, they leveraged a source labeled dataset to enhance training on a target noisy dataset. Initially, they used a mixed-batch training with both training datasets to train a Faster R-CNN model using \textit{BGC} \citep{ros2016training}, ensuring stable gradient directions. They used ADELE method to detect when the network starts memorizing the initial noisy annotations. Next, in the semi-supervised phase, they applied a label correction strategy using test-time augmentation and weighted box fusion techniques to produce confident bounding boxes.

\subsubsubsection{Robust loss function}
\label{Sec:Robust loss function}

Robust loss functions focus mainly on improving the loss to build robust DL network.

To address the problem of model overfitting due to label ambiguity and noisy labels, \citet{sun2022fully} proposed to use \textit{deep log-normal label distribution learning} and \textit{focal loss}. This approach is inspired by \textit{label distribution learning} \citep{geng2016label,gao2017deep}, where an instance is assigned a label distribution, aiming to learn a mapping from instance to label distribution. 
For pneumoconiosis staging on chest radiographs, the authors modeled the label distribution using an asymmetric log-normal distribution.
\begin{equation}
    y^d = \frac{p(y_i | \mu, \delta)}{\sum_{j} p(y_j | \mu, \delta)}
\end{equation}

\begin{equation}
p(y_i | \mu, \delta) = \frac{1}{y_i \sqrt{2\pi} \delta} \exp \left( -\frac{(\log(y_i) - \mu)^2}{2\delta^2} \right)
\end{equation}

$y^d$ is the probability distribution (label distribution) with  $y^d \in [0,1]$. In normal label distribution, the label $y_i$ starts from 1, $\mu$ is the mean value equal to $\log(y_i)$, and $\delta$ is an hyperparameter. 

The KL Divergence loss was employed to enforce label distribution learning by measuring the distance between the label distribution ($y^d$) and the network prediction after the Softmax function. In addition, a regularization term, the cross-entropy loss,  was added to strengthen the learning abilities of the model on unambiguous samples and handle  subjective inconsistencies. The combination of KL divergence loss and  cross-entropy loss forms the focal staging loss.  To resolve optimization inconsistency when using these losses together, an instance-level drop parameter was introduced to skip samples with better predicted results during the optimization process. 

\citet{zhu2021hard} employed the focal loss (Eq.~\ref{Eq:focal_loss}) to improve  training by emphasizing hard samples.
\begin{equation}
 \mathcal{L}_{focal}=-(1-q(y|x))^\gamma \log(q(y|x))
 \label{Eq:focal_loss}
\end{equation}
where $q(y|x)$ is the predicted probability and $\gamma$ is a hyperparameter.
 $\gamma$ was set to 2 to reduce the relative loss for well-classified examples  and focusing more on hard, misclassified one. This technique enhances the robustness of the model against noisy labels and ensures effective learning from challenging data.
 
\citet{hu2021deep} proposed a robust training method,  \textit{Deep Supervised Network with a Self-Adaptive Auxiliary Loss (DSN-SAAL)}, for diagnosing imbalanced CT images.  This framework integrates a novel loss function to address both the effects of data overlap between CT slices and noisy labels. To account for data overlap between CT slices, they adjusted the weight of samples in the Cross-Entropy (CE) loss function.

\begin{equation}
    \mathcal{L}_{CE}=- \sum_{i=1}^c \frac{1- \alpha}{1-\alpha^{k_i}} p(y_i|x) \log (q(y_i|x))
    \label{Equation_CE}
\end{equation}

where $q(y|x)$ is the classifier's output and $p(y|x)$ is the ground-truth label distribution. $k_i$ is the number of samples in the $k^{th}$ class and $\alpha$ is a learnable parameter representing the effective sample factor to measure the ratio of the effective number of samples. 
To combat noisy labels, they introduced the \textit{Reverse Cross Entropy (RCE) loss}:
\begin{equation}
    \mathcal{L}_{RCE}=- \sum_{i=1}^c \frac{1- \alpha}{1-\alpha^{k_i}}q(y_i|x) \log (p(y_i|x)) 
    \label{Equation_RCE}
\end{equation}

Here, $q(y|x)$ is used as the ground truth and $p(y|x)$ is the class probability of the outputs, hence the name reverse cross-entropy.
Finally, the \textit{self-adaptive auxiliary loss} combines the aforementioned losses (Equations \ref{Equation_CE} and \ref{Equation_RCE}) while adding a weighting hyperparameter $\beta$. This approach outperformed the state-of-the-art methods when evaluated on COVID19-Diag and three public COVID-19 diagnosis datasets.

To address annotation subjectivity, \citet{yu2022space} proposed the \textit{Grading Cross Entropy (GCE) loss}, designed to account for the feature continuity of disease grades and  progression of disease grades.  Misclassifications are more likely  between adjacent grades than distant ones. The GCE loss is defined as follows: 
\begin{equation}
    \mathcal{L}_{GCE} = -\sum_{i} p(y_i|x) \log\left(1 - \prod_{j \in \mathbb{N}(i)} (1 - q(y_j|x)) ^{w_{ij}}\right)
\end{equation}
where $p(y_i|x)$ is the $i$-th element of the one-hot encoded label of the input $x$,
$\mathbb{N}(i)$ the neighboring indexes of grade $i$, $q(y_j|x)$ denotes the $j$-th element of the model predictions and $w_{ij}$ represents the weight of grade $j$ with the annotated label $i$. 
This weighting system allows the GCE loss to be more flexible than the CE loss in handling noise by setting different weights to neighboring labels and the annotated label. 

\citet{hermoza2022censor} tackled the problem of predicting survival time from medical images using both censored and uncensored data. They proposed \textit{an Early-Learning Regularization (ELR) loss}, a  regularization loss to manage noisy pseudo labels in survival time prediction. The ELR loss ensures continuous training for samples where the model's prediction aligns with the temporal ensembling momentum (i.e., the ``clean" pseudo-labeled samples) and ceases training for noisy pseudo-labeled samples.
The ELR loss was expressed as follows:
\begin{equation}
    \mathcal{L}_{ELR}(z_i^c)=\log(1- \frac{1}{c} (\sigma (z_i^c)^T \sigma(\tilde{z}_i^c)))
\end{equation}
where $\tilde{z}_i^{c(e)}= \phi \tilde{z}_i^{c(e-1)}+ (1-\phi) z_i^{c(e)}$ is the temporal ensembling momentum of the prediction ($z_i^c$) with $e$ denoting the training epoch and $\phi \in [0,1]$. $\sigma$ is the sigmoid function, $z^c$ is the model's output, and $c$ denotes the number of classes.

\citet{liu2022nvum} proposed \textit{a new training module called Non-Volatile Unbiased Memory (NVUM)}\footnote{\url{https://github.com/FBLADL/NVUM}}, which stores a running average of model logits. They employed a regularization loss to minimize the differences between the model's current logits and those from its initial learning phase. The model $f_\theta$ is trained on a noisy labeled dataset using binary cross-entropy loss, combined with the following regularization term: 
\begin{equation}
    \mathcal{L}_{REG}(\tilde{z}_i^c,z_i^c)= \log (1-\sigma ((\tilde{z}_i^c)^T \sigma(z_i^c)))
\end{equation}
where $\tilde{z}^c$ stores an unbiased multi-label running average of the predicted logits of all training samples and employs the class prior distribution $\pi$ for updating. Initially, $\tilde{z}$ is initialized with zeros and is updated in every epoch as follows:
\begin{equation}
    \tilde{z}_i^{c(e)}=\beta \tilde{z}_{i}^{c(e-1)}+ (1-\beta)({z}_i^{c(e)}-\log \pi)
\end{equation}

where $\beta \in [0,1]$ is a hyperparameter controlling the volatility of the memory storage. $\beta$ was set to $0.9$ representing a non-volatile memory. This regularization enforces consistency between the current model logits and the logits produced at the beginning of the training, assuming robustness to noisy labels in early training. NVUM was evaluated on noisy multi-label imbalanced chest X-ray (CXR) training sets, formed by Chest-Xray14 and CheXpert, and tested  on  clean datasets OpenI and PadChest. The approach outperformed previous state-of-the-art classifiers with mean testing AUC of 0.8865  and 0.8555, respectively.

\citet{shi2020graph} proposed a semi-supervised DL approach, \textit{Graph Temporal Ensembling (GTE)}, which leverages both labeled and unlabeled data while being robust to noisy labels. 
Inspired by Temporal Ensembling (TE) \citep{laine2016temporal}, GTE creates ensemble targets for feature and label predictions  through Exponential Moving Average (EMA) to aggregate feature and label predictions from previous training epochs. Then, the ensemble targets within the same class are aggregated into clusters for further enhancement.The method also utilizes a consistency loss, which  minimizes the discrepancy between the current predictions and the ensemble targets, to form consensus predictions under different configurations. The authors validated the proposed method with extensive experiments on lung and breast cancer datasets, achieving 90.5\% and 89.5\% image classification using 20\% labeled patients on the two datasets, respectively.

\citet{gundel2021robust} addressed the issue of label noise originating from natural language processed medical reports in chest radiography abnormality classification. They measured prior label probabilities  on a subset of training data re-read by 4 board-certified radiologists to enhance  model robustness. These probabilities were used to adjust the weights in the loss function. Sensitivity ($s_{sens}$) and specificity ($s_{spec}$) of the original dataset labels were computed based on the subset of the re-read labels. Sensitivity was defined as   $s_{sens}=\frac{TP}{P}$, and specificity is $s_{spec}=\frac{TN}{N}$  where $TP$ and $TN$ are true positives and true negatives, respectively, based on the re-read subset. $P$ and $N$ are the total number of positive and negative samples in the re-read samples, respectively. To increase the robustness of the model, a regularization term $\mathcal{L}_{noise}$ was added to the binary cross-entropy loss:
\begin{equation}
\begin{split}
\mathcal{L}_{noise} =- \sum_{j =1}^{n} \sum_{i=1}^{c}   [ \lambda_{noise}  [ I_P^{(i)} w_N^{(i)} (1-p(y_i|x_j)) \log q(y_i|x_j) + \\
I_N^{(i)} w_P^{(i)} p(y_i|x_j) \log (1 -q(y_i|x_j)]]
\end{split}
\end{equation}
$I_P$ and $I_N$ are individual regularization weights for positive and negative examples, with  $I_P^{(i)}= 1-s_{sens}^{i}$ 
and  $I_N^{(i)}= 1-s_{spec}^{i}$. $w^{(i)}_P$ and $w^{(i)}_N$ are weight constants to address imbalance, defined as $w^{(i)}_P = \frac{P^{(i)}+N^{(i)}}{P^{(i)}}$ and $w^{(i)}_N= \frac{P^{(i)}+N^{(i)}}{N^{(i)}}$. 
$\lambda_{Noise}$ is a weight parameter controlling the influence of the regularization term. 
In addition, the authors incorporated the correlation between labels observed in chest radiography into the original loss function to further reduce the impact of label noise.

\citet{qiu2023hierarchical} incorporated a regularization loss in their self-training framework, called \textit{Pathin-NL}. This approach used the KL divergence to enforce the similarity between the soft label distribution, estimated using the model's current softmax predictions, and the estimated noise free label distribution,  computed  using the model's softmax output for the previous iteration. They assumed that the majority of images were initially correctly labeled. Thus, the original labels were incorporated into training via standard cross-entropy loss. This prevents the estimated label distribution from deviating significantly from the initial noisy labels. They validated their approach on pathology image classification tasks using glioma and lung cancer datasets from The Cancer Genome Atlas (TCGA). Their method achieved an AUC of 0.872 and 0.977 on the two datasets, respectively.

\subsubsubsection{Graph}
\label{sec:Graph}
Graph-based methods aim to model relationships between images \citep{shi2020graph, ying2023covid} or between patches \citep{xiang2023automatic, yu2022space} in feature space to better detect label noise.

\citet{xiang2023automatic} proposed a weakly supervised model \textit{Graph Convolution Network-Multiple Instance Learning \linebreak (GCN-MIL)} for prostate cancer grading. It consists of: 1) a self-supervised CNN for feature extraction using contrastive loss on unlabeled images, 2) a GCN and attention pooling model for feature aggregation. In the second phase, a graph  was constructed from embedding vectors and their spatial position. DeepGCN convolution was conducted on the graph-structure data to pass information among nodes. Attention pooling over all nodes 
was used for final grading prediction. 
To handle imperfect labels, the model iteratively filtered out noisy samples based on high loss and uncertainty, updating the GCN-MIL model with only clean samples at each iteration.
 
\citet{shi2020graph} proposed a Graph Temporal Ensembling (GTE). The graph-based approach was used to map labeled samples of each class into a cluster. It has shown to be more beneficial for semi-supervised learning than feature consistency which aims to form consensus predictions of feature representations (described in Section~\ref{Sec:Robust loss function}). In contrast, feature consistency has shown significant improvement for combating noisy labels.
 
\citet{yu2022space} presented a framework for pathological cancer grading that addresses space noise (inaccurate boundaries of cancerous areas) and level noise (inaccurate cancer grading).  The framework used a space-aware branch in which  the large image was converted  into a \textit{Multilayer Superpixel (MS) graph},  significantly reducing data size while preserving the global features. These graphs were then processed with a GCN for generating pseudo-masks, which  were then used by the CNN network to fine-tune the binary classification results. For handling level noise, a level-aware branch adopted grouped convolution kernels and a novel grading loss. Finally,  bidirectional cooperation between both branches were conducted, achieving high performances on CAMELYON16, PANDA and HCC datasets with accuracies of 0.9472, 0.7902 and 0.5799, respectively.

\citet{ying2023covid} employed \textit{neighborhood graph regularization} after reducing data dimensionality using PCA. Their aim was to perform manifold learning for ensuring that the reduced-dimensional data retains its original local structure.

\subsubsubsection{Meta-model}
\label{Meta-model}
Few-shot meta-learning aims to train a model that can quickly adapt to a new task using only a few data-points and training iterations \citep{finn2017model}.
To this end, in meta-learning, the model is trained on a set of tasks in a way that the model can quickly adapt to new tasks using only a small number of examples. In the context of DG, meta-learning can tackle the problem of label noise by leveraging the uncertainty of predicted scores and producing meta-models that contain robust features.

\citet{do2021multiple} presented a new \textit{Multiple Meta-model Quantifying (MMQ)}\footnote{\url{https://github.com/aioz-ai/MICCAI21_MMQ}} method designed to enhance medical Visual Question Answering (VQA) by learning meta-annotations and leveraging meaningful features. Their framework includes three modules: 1) \textit{Meta-training} for training a meta-model to extract image features for medical VQA, 2) \textit{Data refinement} which uses auto-annotation to increase training data and manages noisy label by evaluating the uncertainty of predicted score, and 3) \textit{Meta-quantifying} for selecting meta-models whose robust to each others and have high accuracy during the inference phase of model-agnostic tasks.

For meta-training, they followed \textit{Model-Agnostic Meta-Learning (MAML)} \citep{finn2017model}. Considering a model $f$ with its  parameters $\theta$, the updated parameter vector $\theta'$ for a new task $T_i$ with dataset $\{D_i^{tr},D_i^{val}\}$ is given by:
\begin{equation}
    \theta_i'= \theta -	\eta \nabla_\theta \mathcal{L}_{T_i}(f_\theta(D_i^{tr}))
\end{equation}
where $\eta$ is a learning rate.
The model parameters are trained by optimizing for the performance of $f_{\theta_i'}$ with respect to $\theta$ across all tasks. At the end of each iteration, the meta-model parameters are updated using validation sets of all tasks to learn generalized features. Formally, the meta-objective is as follows:

\begin{equation}
\theta= \theta -\beta \nabla_\theta  \sum_{T_i}\mathcal{L}_{T_i}(f_{\theta_i'}(D_i^{val}))
\end{equation}
where $\beta$ is a learning rate.

After meta-training, the meta-models weights are used for data refinement, which aims to enhance the meta-data  by removing samples with  predicted scores below a predefined uncertainty threshold, indicating noisy samples.

The meta-quantifying phase identifies useful meta-models for the medical VQA task by computing a fuse score $S_F$  to quantify performance during the validating process for each meta-model :
\begin{equation}
    S_F=\gamma S_P +(1- \gamma)\sum_{t=1}^k 1-Cosine (z_c^f,z_t^f) \quad \forall z_c^f \neq z_t^f
\end{equation}
where $S_F$ is the fuse score, $\gamma$ is the effectiveness-robustness balancing hyperparameter, $S_P$ is the predicted score over the ground-truth label, and $k$ is the number of candidate meta-models. $z_c^f$ and $z_t^f$  are the feature extracted from the current and the $t$-th meta-model, respectively. Cosine represents the cosine similarity function.

\subsubsection{Collaborative methods}
\label{Sec:Collaborative and iterative methods}
DL methods are prone to overfitting on incorrect labels, which can affect their ability to generalize. To overcome this issue, some approaches have focused on incorporating regularization into the loss function \citep{miyato2018virtual}. However, in some cases, these methods prevent the classifier from achieving optimal performance. On the other hand, some strategies have attempted to estimate the transition matrix \citep{patrini2017making}, a technique that avoids a regularization bias and has the potential to enhance classifier performance. However, accurately estimating transition matrix is challenging, in particular with datasets that are imbalanced. A promising solution to avoid the complexities of estimating the noise transition matrix involves focusing on training with a subset of carefully selected samples. This approach aims to filter out clean instances from the noisy data for network training. In this context, collaborative methods via training two or more models leverages the cooperation between models for improving the performances of DL models. These methods encompass co-training, co-teaching, and knowledge distillation.  

\subsubsubsection{Co-training}
\label{Sec:Co-training}
Co-training is a machine learning technique where two or more models are trained separately on distinct views of the data, and their predictions are used to enhance each other's learning process.

\citet{zhou2023combating} proposed a co-training approach to tune a single target network for disease classification. They pre-trained multiple reference networks to handle label uncertainty. To co-optimize the target network, they introduced a \textit{Disentangled Distribution Learning (DDL)} strategy, which disentangle the multiple reference models' predictions into a hard \textit{Majority Confident Label (MCL)} vector (a pseudo cleaned ground-truth) and a soft \textit{Description Degree Score (DDS)} vector.
The MCL vector was computed by counting the number of networks giving positive and negative predictions for the corresponding disease label. The DDS vector was computed using the average over all the predictions. To optimize the target network, they used KL divergence based on the confidence-weighted relative entropy of the hard majority label vector with respect to the predictions of the target network. Moreover, they proposed \textit{inter- and intra-instance consistency regularization} to enforce the target network to provide consistent predictions for images with similar medical findings. This involved using KNN smoothing modules and image augmentation. $K$ nearest neighbors of an image (the anchor image) were computed based on the fixed soft label distribution. Then, the target network was constrained to produce similar predictions for the anchor image and its $K$ nearest neighbors. In addition, the anchor image was also augmented into different views and the target network was constrained to have the same prediction for these views. Experiments performed on chest X-ray and fundus image dataset, showed that the proposed approach is outperforming state-of-the-art methods.

\citet{xue2022robust} proposed a \textit{co-training with global and local representation learning framework}. Two independent teacher-student networks   were trained with different image augmentation and initialization strategies to ensure distinct weight parameters. After one epoch of training, a Noisy Label Filter (NLF) divided the data into clean and noisy samples based on the teacher encoder's predictions. Specifically, the NLF used a two-component Gaussian Mixture Model (GMM) to fit the max-normalized cross-entropy loss of the training data via the Expectation-Maximization algorithm. Clean and noisy samples were then crossly sent to the peer networks. Rather than removing the noisy labeled sample,  a \textit{self-supervised learning strategy} was proposed.  \textit{A local contrastive loss} was applied on noisy samples, encouraging the network to learn robust representations by minimizing differences between augmented views of the same image and maximizing differences from other images. A \textit{global relation loss} was applied to align the inter-sample relationship of samples between the teacher and student model.
Experiments on datasets such as Histopathologic Lymph Node, ISIC Melanoma, Gleason 2019, and CXP showed that this approach consistently outperformed other state-of-the-art methods, especially in scenarios with high noise ratios.

\subsubsubsection{Co-teaching}
\label{Sec:Co-teaching}
Co-teaching is a paradigm where two models are jointly trained with each model selecting the instances to train the other model. Since each model is initialized differently, each model learns a different decision boundary, resulting in different selection of training instances.

\citet{zhu2023robust} presented a robust \textit{co-teaching} paradigm that cross-trains two DL networks simultaneously to select small-loss samples for training. Their approach comprises two modules. First, an \textit{Adaptive Noise Rate Estimation} module was employed for estimating the dataset's noise rate by using the maximum validation accuracy from the networks. This noise rate was used to set the percentage of small-loss samples selected as probably clean samples in the subsequent module. Second,  a \textit{Consistency-based Noisy Label Correction} module was applied to select probably clean samples based on their loss ranking according to both networks and to relabel highly suspected noisy samples (samples with consistent predictions and high confidence) using consistent predictions. The corrected samples were aggregated with small-loss samples into ``a corrected set", which was used for training the network in the next iteration. This approach showed promising performance when tested on public skin lesion datasets (ISIC-2017, and ISIC-20019) and a constructed thyroid ultrasound image dataset. 

One drawback of co-teaching is that ordering data based on their loss may overlook difficult examples that may be correctly labeled but hard to train. To overcome this issue, \citet{peng2020noise} proposed co-weighting, which trains two DL networks simultaneously, teaching each other with every mini-batch. Unlike co-teaching,  co-weighting dynamically re-weights samples of the current batch.  Noisy samples are identified and excluded by analyzing the statistical features of predictive history and only hard informative samples are retained. In this approach, the prediction history stores learning events that correspond to increases in predictions between consecutive updates and forgetting events that correspond to  decreases in predictions. Noisy samples, identified by frequent forgetting events, are excluded. The noise ratio was estimated using noisy cross-validation \citep{chen2019understanding}. The reserved samples underwent a ranking process. Experiments on DigestPath2019 and the colorectal tumor dataset showed high average accuracy ($>0.915$) in 5-fold cross-validation, outperforming co-teaching.

\citet{zhu2021hard} also proposed an improvement over co-teaching framework. They proposed a \textit{hard sample aware noise robust learning method}\footnote{\url{https://github.com/bupt-ai-cz/HSA-NRL/}}, composed of two phases: a \textit{label correction} phase and a \textit{Noise Suppressing and Hard Enhancing} (NSHE) phase. The label correction phase produces an ``almost clean dataset" by pre-discarding most of the noisy samples using a self-training strategy (described in Section~\ref{sec: Re-labeling}). The almost clean dataset is then used in the NSHE phase, which enhances hard samples while suppressing the remaining noisy ones through a colearning architecture. Two DL networks, $f_1(x,\theta_1)$ and $f_2(x,\theta_2)$, are initialized with the same backbone and parameters. Inspired by MoCo \citep{he2020momentum}, the parameters of the first DL network $\theta_1$ are updated by back-propagation, while the parameters of the second DL network $\theta_2$ are updated using a momentum-based approach:
\begin{equation}
    \theta_2 \leftarrow m\theta_2 +(1-m)\theta_1
\end{equation}
where $m\in [0,1)$ denotes a momentum coefficient. $\theta_2$ evolve more smoothly than $\theta_1$. At each epoch, $f_2$ selects training data for $f_1$ by ranking samples according to their prediction values, excluding those with small prediction probabilities from back-propagation. 

\citet{liu2021co} proposed a \textit{Co-correcting} strategy\footnote{\url{https://github.com/JiarunLiu/Co-Correcting}} to address noisy labels by simultaneously training two DL networks with identical architecture. The parameters are updated using an “updated by agreement" principle,  assuming that instances with small losses are clean and collecting their gradients when agreement occurs. The framework consists of three modules. A dual-network module based on mutual learning, where networks are trained by selecting clean samples based on small losses and mutual agreement. A curriculum learning module, in which co-correcting introduces a label correction strategy by increasing difficulty from easy tasks to harder ones.  Finally,  a label updating module based on a probabilistic estimation of whether the label is noise-free (label distribution) similar to the PENCIL framework \cite{yi2019probabilistic}. The idea is to update both network parameters and label estimations as label distributions. The estimated label distribution serves as a pseudo-label. It is initialized based on the noisy labels and continuously updated using backpropagation. 

\subsubsubsection{Knowledge distillation}
\label{sec:Knowledge distillation}
\textit{Knowledge distillation} involves transferring knowledge from one model to the other. A student model trained on noisy labels is guided by a teacher model. Initially, a teacher model is trained on a clean dataset. In parallel, a student model is trained using a combination of original noisy labels and the teacher's output predictions, which serves as pseudo labels. As training progresses, better guidance is provided to the student model since the prediction of the teacher model becomes more reliable.

\citet{li2021bootstrap} proposed a novel \textit{Bootstrap Knowledge Distillation (BKD) method} to gradually improve label quality and reduce noise. The method was applied for lung disease classification using the CheXpert  and  Chest X-ray14 datasets. For the CheXpert dataset, the teacher model was trained on certain data. Then, various strategies were employed to train it on the entire dataset. For handling uncertain labels,  three different strategies were adopted: mapping all uncertain labels to 0, to 1, or to the output probability of an auxiliary model. 
The third strategy outperformed a baseline CNN with an ensemble of 30 checkpoints. For Chest X-ray14 dataset, they used a pre-trained model to select a clean subset. Samples with an output probability larger than 0.55 and a ground truth 1 were considered certain positive, while those with an output probability smaller than 0.45 and ground truth label 0 were certain negatives; the rest were uncertain labels. Their method showed good performances, outperforming state-of-the-art methods on most pathologies.

\section{Public medical datasets for generalization research}
\label{sec: Public medical datasets}
There exist many public datasets which have been adopted for generalization research in the medical field. For example, to prevent domain shifting, some mutlti-institutional datasets have been proposed for segmentation problems  \citep{menze2014multimodal,campello2020multi} and image  reconstruction \citep{huang2022evaluation} \footnote{\url{https://brain-development.org/ixi-dataset/},\url{https://www.kaggle.com/c/second-annual-data-science-bowl/data},\url{http://mridata.org/}}. In this section, we present the publicly available datasets that were used for classification experiments in the selected articles. Table~\ref{tab:public_datasets_1} summarizes these public medical datasets which can be used for generalization research.
For a more comprehensive understanding for readers, we will give brief details for public datasets available as part of a challenge.

{\textbf{MIDOG datasets}} Mitosis Domain Generalization (MIDOG) dataset targets the detection of mitotic figures in histopathology images under domain shift regime.
\begin{itemize}
\item MIDOG 2021 dataset~\citep{aubreville2023mitosis}: This dataset was part of the MICCAI MIDOG 2021 challenge which aims to evaluate methods that mitigate domain shift and derive scanner-agnostic algorithms. It addresses DG in histopathology. The main task was mitosis detection in breast cancer. The challenge dataset features 300 cases, 6 scanners, and more than 2500 mitosis. The domains are defined by scanner types.
\end{itemize}
{\textbf{CAMELYON datasets}} Cancer Metastases in Lymph nodes challenge (CAMELYON) datasets target the automated detection of cancer metastases in Whole-Slide Images (WSIs) of sentinel lymph nodes.
\begin{itemize}
    \item CAMELYON16 dataset~\citep{bejnordi2017diagnostic} originates from CAMELYON16, in 2016. The dataset includes 399 WSIs collected from 2 centers.
    \item PatchCamelyon~\citep{veeling2018rotation} is a large-scale patch-level dataset derived from Camelyon16 dataset.
    \item CAMELYON17 dataset~\citep{koh2021wilds} originates from the CAMELYON17 challenge which was held in 2017. In comparison to CAMELYON16 which focuses on slide level analysis, CAMELYON17 focus on patient level analysis. The dataset comprises 1000 WSIs collected from 5 centers.
\end{itemize}

{\textbf{LUNA-16}}: The goal of LUNA-16 challenge is the  automated detection of pulmonary nodules in thoracic Computed Tomography (CT) scans \citep{setio2017validation}. This challenge use data from a large public LIDC-IDRI dataset ~\citep{armato2011lung}. More precisely, scans with a slice thickness greater than 2.5 mm were excluded. The resulting dataset contains 888 CT scans.

{\textbf{PANDA dataset}}: The goal of Prostate Cancer Grade Assessment (PANDA) challenge \citep{bulten2022artificial} is the diagnosis of prostate cancer in biopsies. It aims to develop AI algorithms for Gleanson grading. In total, the PANDA dataset comprises 12,625 WSIs of prostate biopsies  retrospectively collected from 6 different sites for algorithm development, tuning and independent validation.
Cases for development, tuning and internal validation originated from two European (EU) centers: Radboud University Medical Center, Nijmegen, the Netherlands and Karolinska Institutet, Stockholm, Sweden. The external validation data consisted of a US (741 cases) and an EU set (330 cases).

\begin{table*}[!htpb]
\begin{center}
\adjustbox{max width=\textwidth}{%
    \begin{tabular}{ccccc}
    \toprule
Dataset &   Modality (Organs) & Number of cases & Reference \\ 
    \toprule
MIDOG 2021 dataset~\citep{aubreville2023mitosis} &  \multirow{7}{*}{Histopathology (Breast)} &  300 images &\url{https://midog2021.grand-challenge.org/}\\
VGH \citep{beck2011systematic}&  &5,920 images& \multirow{2}{*}{\url{https://tma.im/tma_portal/C-Path/supp.html}}\\
 NKI  \citep{beck2011systematic}& & 8,337 images&  \\
 Camelyon16 WILDS dataset~\citep{bejnordi2017diagnostic}&  & 399 WSI &  \url{https://camelyon16.grand-challenge.org/Data/}  \\ 
PatchCamelyon \citep{veeling2018rotation} & & 327,680 color images & \url{https://patchcamelyon.grand-challenge.org/}\\
Camelyon17 WILDS dataset~\citep{koh2021wilds}&  & 1000 WSI &  \url{https://camelyon17.grand-challenge.org/Data/} \\ 
TCGA-BRCA &&1,098 cases&  \url{https://portal.gdc.cancer.gov/projects/TCGA-BRCA}\\ 
\midrule
BACH dataset& Microscopy,Histopathology (Breast)&400 microscopy images 30 WSI& \url{https://iciar2018-challenge.grand-challenge.org/Dataset/} \\ 
\midrule
UBC-OCEAN &  Histopathology (Ovaries) &  538 training images & \url{https://www.kaggle.com/competitions/UBC-OCEAN/data} \\
\midrule
PANDA& \multirow{3}{*}{Histopathology (Prostate)}& 12,625 WSIs& \url{https://www.kaggle.com/c/prostate-cancer-grade-assessment} \\
DiagSet-B& &  4675 scans & \url{https://github.com/michalkoziarski/DiagSet} \\
SICAPv2 \citep{silva2020going}&   & 155 biopsies (95 patients)& \url{https://data.mendeley.com/datasets/9xxm58dvs3/1} \\
\midrule
TUPAC-16\citep{bertram2020pathologist} &Histopathology (Colon)&1076 cases& \url{https://tupac.grand-challenge.org/Dataset/}\\
\midrule
 Kather16 \citep{kather2016multi}  &\multirow{10}{*}{Histopathology (Colon)} &  5,000 patches & \url{https://zenodo.org/records/53169}  \\
Kather19 \citep{kather2019predicting} &  &100,000 patches &  \url{http://dx.doi.org/10.5281/zenodo.1214456} \\
 CRC-TP \citep{javed2020cellular} &   &196,000 patches &  \url{https://warwick.ac.uk/TIAlab/data/crchistolabelednucleihe/.}
 \\ 
  IHC \citep{linder2012identification} &  &1,376 
 images & \url{http://fimm.webmicroscope.net/supplements/epistroma}\\ 
 CRC-VALHE-7K  & & 7,180 image patches &  \url{https://zenodo.org/records/1214456}\\ 
 Stanford-CRC \citep{yamashita2021deep} &    &  66,578 image tiles & \url{https://github.com/rikiyay/MSINet} \\
CRC-DX-TRAIN dataset &   & 93,408 image tiles & \url{https://zenodo.org/records/2530835#.XwCkDZNKhTY} \\
CRC-DX-TEST dataset& &99,904 image tiles &
\url{https://zenodo.org/records/2530835#.XwCkDZNKhTY} \\
Chaoyang Dataset \citep{zhu2021hard}& & 6,160 WSI&  \url{https://bupt-ai-cz.github.io/HSA-NRL/}\\
DigestPath2019 \citep{li2019signet}& & 690 patients& \url{https://digestpath2019.grand-challenge.org/} \\
\midrule 
TCGA-LGG& \multirow{2}{*}{Histopathology (Brain)} & 516 cases& \url{https://portal.gdc.cancer.gov/projects/TCGA-LGG} \\
TCGA-GBM& & 617 cases& \url{https://portal.gdc.cancer.gov/projects/TCGA-LGG} \\ 
\midrule
TCGA-LUAD& \multirow{2}{*}{Histopathology (Lung)} & 585 cases& \url{https://portal.gdc.cancer.gov/projects/TCGA-LUAD} \\
TCGA-LUSC& &504 cases& \url{https://portal.gdc.cancer.gov/projects/TCGA-LUSC} \\ 
\midrule
ADNI-1 \citep{jack2008alzheimer} & \multirow{3}{*}{MRIs (Brain)} &  748 subjects& \url{https://adni.loni.usc.edu/}\\
ADNI-2 \citep{jack2008alzheimer} && 708 subjects &\url{https://adni.loni.usc.edu/} \\
AIBL \citep{ellis2009australian} & &549 subjects&  \url{https://adni.loni.usc.edu/}\ \\ \midrule
ISIC-2017 \citep{codella2018skin} & \multirow{6}{*}{Dermoscopic images} & 2,750  images & \multirow{2}{*}{\url{https://challenge.isic-archive.com/data/}}\\
ISIC-2019\citep{tschandl2018ham10000} & &33,569 images& \\
HAM10000 (HAM)\citep{tschandl2018ham10000} & &10,000 images& \url{https://dataverse.harvard.edu/dataset.xhtml?persistentId=doi:10.7910/DVN/DBW86T} \\
Dermofit (DMF)& &1,300images& \url{https://homepages.inf.ed.ac.uk/rbf/DERMOFIT/datasets.htm}\\
Derm7pt (D7P) \citep{kawahara2018seven} & &2,000 images& \url{http://derm.cs.sfu.ca} \\
\midrule
INbreast \citep{moreira2012inbreast}&  \multirow{3}{*}{Mammography (Breast)}& 115 cases& \url{https://www.kaggle.com/datasets/ramanathansp20/inbreast-dataset} \\
OPTIMAM dataset \citep{halling2020optimam} & & 179,326 cases &  \url{https://medphys.royalsurrey.nhs.uk/omidb/} \\
BCDR \citep{lopez2012bcdr} & &3,703 digitised film mammograms&  \url{https://www.medicmind.tech/cancer-imaging-data}\\
\midrule
LIDC-IDRI~\citep{armato2011lung}  &\multirow{4}{*}{CT(Lung)}&  1,018 scans (1,010  subjects)& \url{https://wiki.cancerimagingarchive.net/pages/viewpage.action?pageId=1966254}\\
LUNA-16 & & 888 scans &  \url{https://luna16.grand-challenge.org/Data/}\\
LUNA-DG \citep{yin2022afa}& &  887 scans  &\url{https://github.com/meisun1207/LUNA-DG} \\
NLST  \citep{national2011national,national2011reduced} && 25,681 patients (77,040 images)& \url{https://cdas.cancer.gov/nlst/}  \\ 
COVID19-Diag& & 226 CT volumes & \url{https://github.com/MLMIP/COVID19-Diag} \\
\midrule
ChestX-ray14 (NIH)~\citep{wang2017chestx}  & \multirow{7}{*}{X-ray (Chest)} & 112,120 images (30,805 subjects) & \url{https://nihcc.app.box.com/v/ChestXray-NIHCC}  \\
CheXpert (CXP)~\citep{irvin2019chexpert}  &&224,316 images (65,240 subjects) & \url{https://stanfordmlgroup.github.io/competitions/chexpert/}
\\
MIMIC-CXR (MMC)~\citep{johnson2019mimic} & &  377,110 images (65,179 subjects) & \url{https://physionet.org/content/mimic-cxr/2.0.0/} \\
PadChest~\citep{bustos2020padchest}  & &160,000 images(67,000 subjects) &\url{https://bimcv.cipf.es/bimcv-projects/padchest/} \\
Open-i dataset \citep{demner2016preparing} &&8,121 images &\url{http://openi.nlm.nih.gov/} \\
Tawsifur \citep{rahman2021exploring,chowdhury2020can} && 931 images & \url{https://www.kaggle.com/datasets/tawsifurrahman/covid19-radiography-database} \\
Skytells & &1,017 images&  \url{https://github.com/skytells-research/COVID-19-XRay-Dataset}\\
 \bottomrule
 UK Biobank retinal photography dataset \citep{sudlow2015uk} &\multirow{11}{*}{Fundus photography (Eye)}  & 58,700 patients & \url{https://www.ukbiobank.ac.uk/}\\ 
EyePACs& & 88,702 images & \url{https://www.kaggle.com/c/diabetic-retinopathy-detection/data} \\
APTOS& & 3,662 images&  \url{https://www.kaggle.com/c/aptos2019-blindness-detection}\\
Messidor& &1,200 images& \url{https://www.adcis.net/en/third-party/messidor/} \\
PALM& & 1,200 images&  \url{https://palm.grand-challenge.org/}\\ 
AV-DRIVE \citep{hu2013automated}& &40 Images&  \url{https://drive.grand-challenge.org/}\\ 
LES-AV& &22 images& \url{https://figshare.com/articles/dataset/LES-AV_dataset/11857698} \\
HRF& & 45 images& \url{https://www5.cs.fau.de/research/data/fundus-images/} \\ 
  REFUGE & &  1,200 images& \url{https://refuge.grand-challenge.org/} \\
   REFUGE2& &   2,000 images& \url{https://refuge.grand-challenge.org/} \\
  STARE& & 20 images& \url{https://cecas.clemson.edu/~ahoover/stare/probing/index.html} \\
   RIGA & &750 images&  \url{https://academictorrents.com/details/eb9dd9216a1c9a622250ad70a400204e7531196d}\\ 
   DDR  & & 13,673 images&  \url{https://drive.google.com/drive/folders/1z6tSFmxW_aNayUqVxx6h6bY4kwGzUTEC}\\
   RIMONEv2 \citep{batista2020rim}& &455 images & \url{https://medimrg.webs.ull.es/}\\
FGADR & & 1,842 images& \url{https://csyizhou.github.io/FGADR/} \\ 
\midrule
Endovis Challenge dataset \citep{allan20192017}& Endoscopic (Abdominal organs) &  8 sequences & \url{https://endovissub2017-roboticinstrumentsegmentation.grand-challenge.org/} \\ \midrule
 Heidelberg colorectal dataset \citep{maier2021heidelberg}&Laparoscopy (Colorectum) & 30 laparoscopic videos& \url{https://robustmis2019.grand-challenge.org/Data/} \\
 \midrule
 CholecSeg8k \citep{hong2020cholecseg8k}& Laparoscopy (Abdomen) & 17 videos from Cholec80 dataset  &  \url{http://camma.u-strasbg.fr/datasets}\\  \midrule
 SurgicalActions \citep{schoeffmann2018video} &Laparoscopy (Gynecologic organs) &160 videos & \url{http://ftp.itec.aau.at/datasets/SurgicalActions160/} \\ \midrule
Cataract-101 \citep{schoeffmann2018cataract}& Video (Eye)&101 cataract surgeries &  \url{http://ftp.itec.aau.at/datasets/ovid/cat-101/}\\
 \midrule
   PathVQA \citep{he2020pathvqa}& Multiple modalities (multi-organs) &  4,998 pathology images (multi-organs)& \url{https://github.com/UCSD-AI4H/PathVQA} \\
   VQA-RAD\citep{lau2018dataset}& Radiology (multi-organs)& 315 radiology images.& \url{https://osf.io/89kps/} \\
   \midrule
   OrganCMNIST&CT (Abdomen)& 23,583&  \url{https://medmnist.com/}\\
    \bottomrule
    \end{tabular}}
    
    \caption{Public medical datasets used for generalization research. All hyperlinks in this paper were retrieved on 22 March 2024.}
    \label{tab:public_datasets_1}
\end{center}
\end{table*}

\section{Discussion}
\label{sec: Discussion}

One of the ultimate goals of DL models in healthcare is to achieve good generalization performances for wider deployment. This desideratum is of critical importance for DL models to be employed in the real world.
However, domain shift is almost inevitable in the medical field. Medical data is heterogeneous, exhibiting significant variability due to diverse imaging modalities, patients demographics, and disease characteristics. These factors are responsible for the occurence of  \textit{covariate shift}. Besides, data is typically collected in diverse scenarios (e.g., mass screening, city consultations, hospital appointments, etc.), possibly in different countries, implying different annotation guidelines, levels of expertise, etc. These factors lead to the manifestation of \textit{concept shift}. For these reasons, we suspect domain shifts are particularly pronounced in the medical domain, compared to general-purpose computer vision tasks, where imaging devices (typically cameras) are more homogeneous and concepts (animal species, building types, etc.) are more universal. Facing this domain shift, it is crucial to ensure that DL will perform robustly, reliably and fairly when making predictions about data different from the training data.
In this paper, we have presented state-of-the-art strategies for the development of generalized method for medical image classification. Depending on the type of shift, two main categories of methods were identified: covariate shift-based methods and concept shift-based methods.

Hereafter, we will first present our analysis of the methods and examine the recent trends in DG development (Section~\ref{sec:SOTA medical image classification for generalization research in the literature}). Next, we will discuss connections with other research fields (federated learning, fair AI, causal AI, etc.) (Section~\ref{sec:Related research to DG}). Afterward, we will address practical issues in implementing and evaluating a DG method (libraries, evaluation strategies, etc.) (Section~\ref{sec:Implementation}). Finally, we will conclude with future directions and promises (subpopulation shift, open DG, etc.) (Section~\ref{sec:Future challenges}), and limitations (Section~\ref{sec: Limitations}).

\subsection{What are the state-of-the-art methods in medical image classification for generalization research in the literature? }
\label{sec:SOTA medical image classification for generalization research in the literature}
Our present taxonomy is based on the analysis of existing studies in generalization research in medical images. For covariate shift, we identified three main categories: data manipulation, representation learning, and learning methods. For concept shift, our taxonomy proposes three main categories: data adjustment and transformation, learning methods, and collaborative methods. In this section, we will first start with a critical analysis of the current state of the field and we will draw conclusions on the benefits of current DG methods, through the analysis of their results on challenge data (Section~\ref{sec:Lessons learnt from challenges}). We will then examine the recent trends in DG development (Section~\ref{sec:Trends in DG}).

\subsubsection{Lessons learnt from challenges}
\label{sec:Lessons learnt from challenges}
Our survey has shown that the generalization problem is common for a variety of modalities: histopathology, X-ray, fundus photographs, ultrasound, etc.

The authors of the reviewed studies used different protocols and datasets splits, making it difficult to compare the results effectively. For a comprehensive comparison, we have included the results of the methods proposed in the MIDOG challenge (Table \ref{tab:Results_MIDOG}).

\begin{table*}[!htpb]
\begin{center}
\adjustbox{max width=\textwidth}{%
    \begin{tabular}{ccccccc}
    \toprule
Article  &   Category& Subcategory  &  F1 score on the preliminary test set& F1 score on the final test set & Other test set  \\ 
    \toprule
\citet{kurian2022domain}& \multirow{6}{*}{Data Manipulation} & Data homogenization & 0.0030  & -- &--\\
 \cline{3-6} 
\citet{li2022single}&  & \multirow{5}{*}{Data augmentation}    &   -- & --  &Average accuracy  0.6285  \\
\citet{chung2021domain}& & &\textbf{0.7548}&\textbf{0.7243}& -- \\
\citet{dexl2021mitodet}&  && 0.7138&0.6963&--  \\
\citet{lafarge2021rotation}&  & & 0.6828&0.6319&-- \\
\citet{long2021domain} & & &   0.7500&0.7010& --\\ \hline
\citet{wilm2021domain} &  Representation learning &Adversarial &  0.750&  0.7183 & -- \\\hline 
\citet{razavi2021cascade} &  Learning strategies &  Multi-task learning    &   0.7492 &0.7064& --\\
    \bottomrule
    \end{tabular}}
    \caption{Generalizing Methods proposed for the  MIDOG challenge.}
    \label{tab:Results_MIDOG}
\end{center}
\end{table*}

A total of 17 methods were submitted for the MIDOG challenge~\citep{aubreville2023mitosis} final test. These methods were compared to a reference DG approach \citep{wilm2021domain}, presented in Section~\ref{sec:Adversarial} (penultimate row in Table\ref{tab:Results_MIDOG}). This approach reduces covariate shift in the feature space by using adversarial training, belonging to “Representation learning-- Adversarial category" in our taxonomy. In addition, a non-DG baseline was considered, named CNN baseline, with the same network topology as the reference approach but only trained using standard image augmentation. Among the submitted methods to the final phase, four methods \citep{chung2021domain,dexl2021mitodet, lafarge2021rotation,long2021domain} were described in this survey and belong to the “Data manipulation--Data augmentation" category (Section~\ref{sec:Data augmentation}),  and one method \citet{razavi2021cascade}  belongs to the “Learning strategies-- Multi-task learning" category.

In contrast to the aforementioned methods, \citet{li2022single} considered single DG setting using the MIDOG dataset. They trained their model using data coming from one scanner domain and then evaluated it on all the other domains. This process was repeated with each domain serving as the training data. Finally, they have computed the mean performance across all unseen domains.

The findings from the MIDOG 2021 competition suggest that through the use of effective augmentation techniques and sophisticated DL architecture models, domain shift between different whole slide imaging scanners can be addressed to some extent. Despite promising overall performances, the results on unseen Scanners were considerably weaker, indicating that domain shift is not completely covered by the algorithms. This highlights that the problem of DG is not solved yet: there is a need for developing more robust algorithms.

In the same context of histopathology, a more recent challenge UBC-OCEAN\footnote{\url{https://www.kaggle.com/competitions/UBC-OCEAN}}  aimed to classify ovarian cancer subtypes based on histopathology images. Owkin's team has won the competition\footnote{\url{https://www.kaggle.com/code/jbschiratti/winning-submission}}. Their solution consisted of using Phikon, Owkin's foundation model for digital pathology. It is a self-supervised foundation model \citep{filiot2023scaling}, which consists of a ViT-Base pre-trained with iBOT on 40 million tiles from the TCGA dataset. Specifically, they trained an ensemble of Chowder \citep{courtiol2018classification} models (multiple instance learning models) on top of Phikon tile embeddings. Then, they used high entropy predictions to detect outliers. These results suggest that the development of foundation models in the medical field pave the way for improving the generalizability of DL models. In particular, the pre-training strategies are important for enhancing the performances of DL models.

\subsubsection{ Trends in DG}
\label{sec:Trends in DG}

\begin{figure*}[ht!]
    \centering
    \begin{tabular}{cc}
\includegraphics[width=9cm,height=6cm]{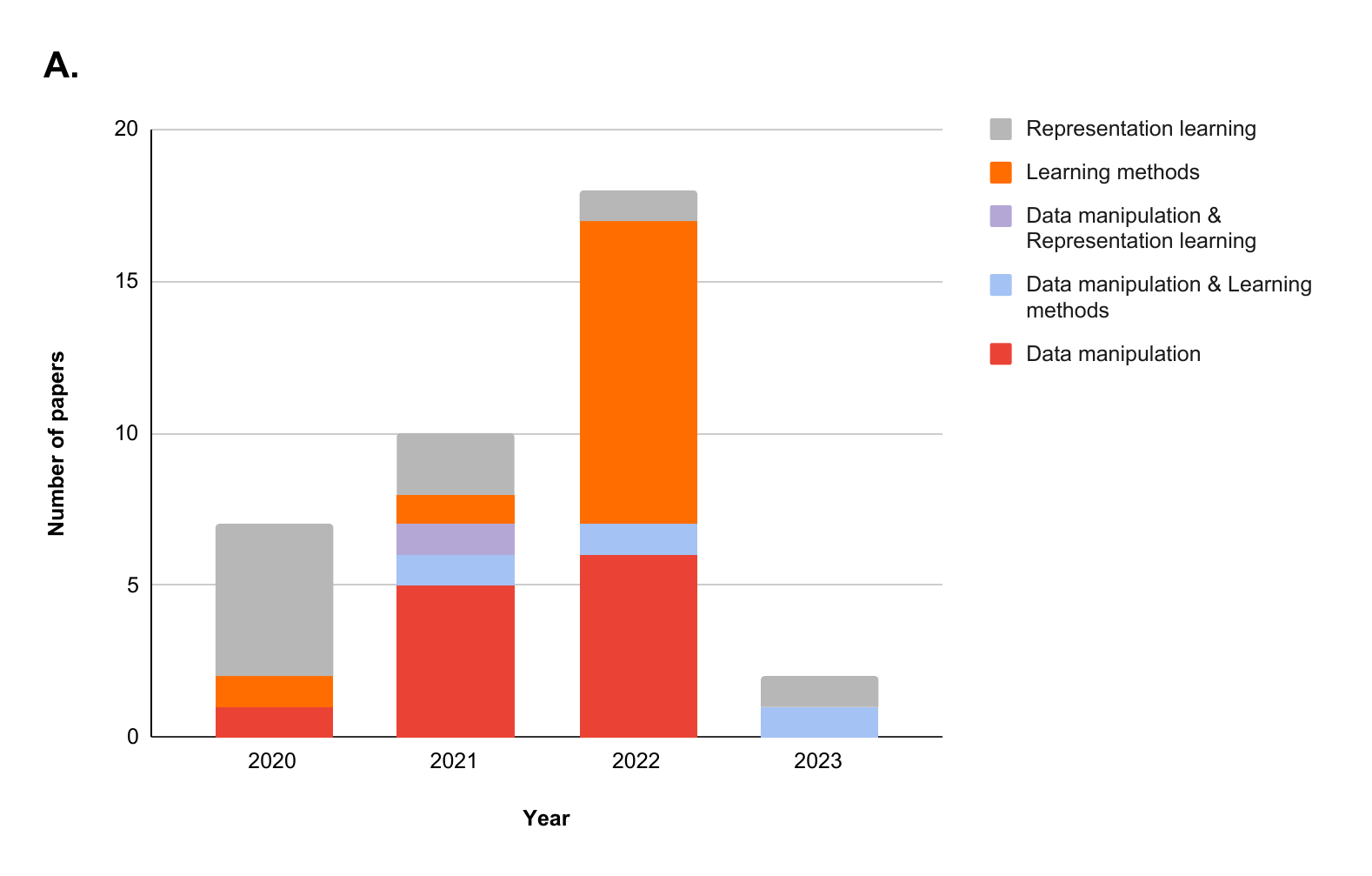} &
\includegraphics[width=9cm,height=6cm]{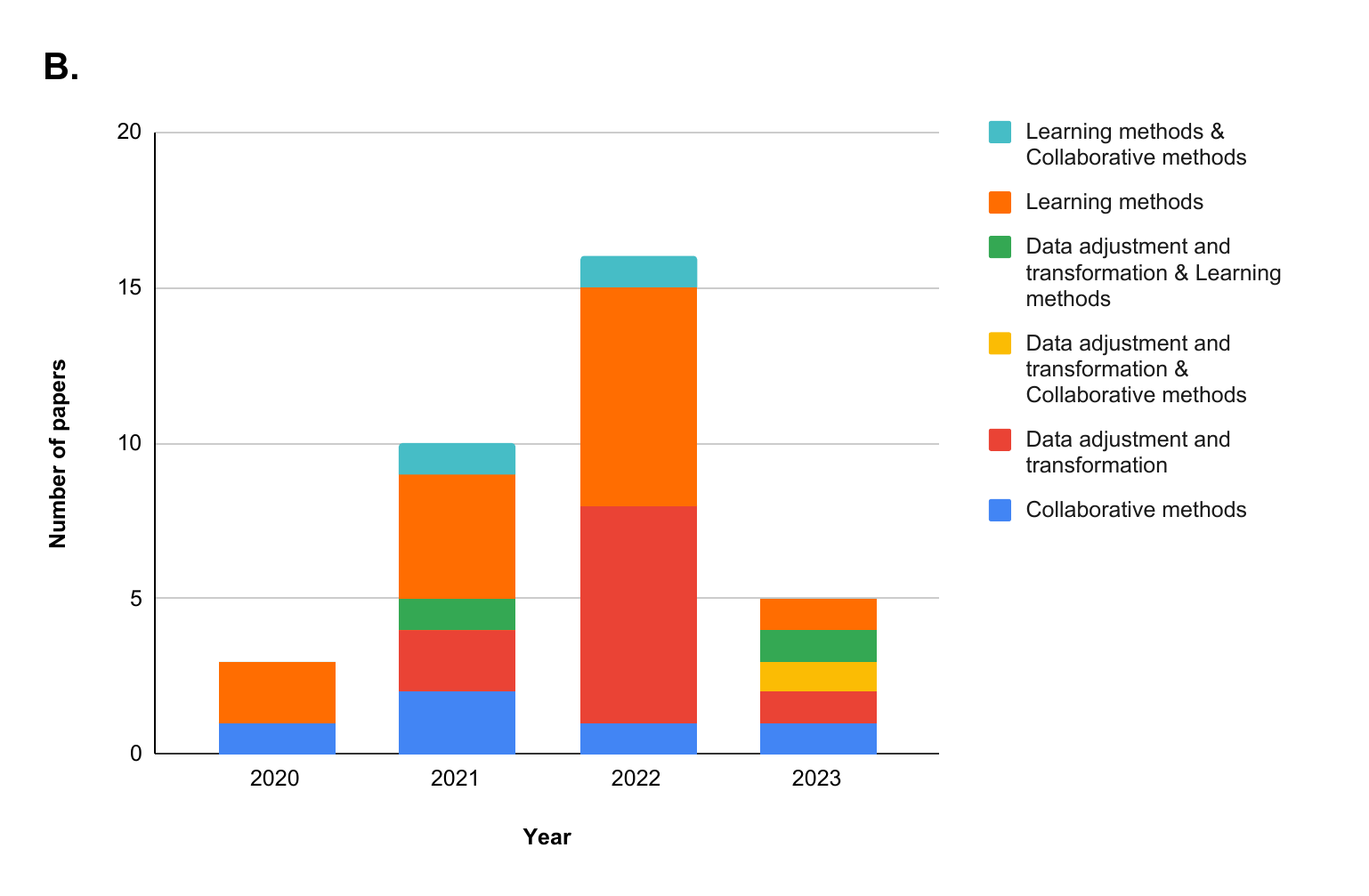} \\
    \end{tabular}
    \caption{Number of paper per year for covariate shift methods (A) and concept shift methods (B). Publications up to April 2023. A unique category is created for papers belonging to multiple categories simultaneously. }
    \label{fig: Figure_6}
\end{figure*}

Figure~\ref{fig: Figure_6} shows the number of papers per category for methods dealing with covariate shift (Figure~\ref{fig: Figure_6} A) and methods with concept shift (Figure~\ref{fig: Figure_6} B). In both graphs, the number of papers tends to increase over the years, suggesting that the generalization research in the medical field is emerging (the decrease in 2023 simply indicates that the reviewed period ends in April 2023). For the methods dealing with covariate shift, it can be seen that learning based methods are showing a significant increase over the years. Data manipulation based methods are showing a smooth increased evolution. On the other hand, we note a slight decreasing evolution for representation learning method.

For methods dealing with concept shift, we also noted that the number of papers employing learning methods is increasing through the years. 
We note that some methods use two categories simultaneously (i.e., “data manipulation" and “representation learning", ``data adjustment and transformation" and “learning methods").This suggests that combining methods could be also studied in future research to enhance the results.  
 
Indeed, learning methods, more precisely based on self-supervised learning, are becoming more and more prominent in the field of generalization research. In this context, a promising avenue for DG is the development of foundation models, a large AI model developed using a massive amount of unlabeled data on a large scale, that can be customized for a wide range downstream tasks.

For example, in the field of ophthalmology, a foundation model for retinal images, RETFound \citep{zhou2023foundation}, was proposed. 
It underwent pre-training on 1.6 million unlabeled retinal images through self-supervised learning, leveraging the weights from a model that was trained on natural images from ImageNet using the same self-supervised learning strategy.
RETFound has achieved increased generalization performances in the diagnosis and prognosis of sight-threatening eye diseases on unseen datasets, when compared to other pretraining strategies. These pretraining strategies used the same model architecture and the fine-tuning process but only differed with the pretraining process. For instance, one classical pretraining strategy consisted of pretraining the model on natural images by means of supervised learning. Other more sophisticated strategies employed self-supervised learning pretraining scheme using either natural images or retinal images.

In histopathology, many foundation models have been proposed \citep{alfasly2024foundation,lai2023domain,filiot2023scaling,yellapragada2024pathldm}. Among these models, \citep{filiot2023scaling} was adopted for the UBC-OCEAN competition and was the winner. 

\subsection{What are the related areas the generalization research?}
\label{sec:Related research to DG}
This section presents the link between generalization research and other learning methods such as federated learning, fairness, and causality. 

\subsubsection{Federated learning}
\label{sec: Federated learning}
Several algorithms \citep{shen2022cd2, andreux2020siloed, li2016revisiting} described in this paper used Federated Learning (FL) in their framework. FL, notably, can be regarded as one of the most practical application of DG in medical imaging. The distributed, heterogeneous data in FL, renders it an appealing scenario for implementing DG in FL applications.
In this context, the medical data is distributed across multiple domains (organizations/hospitals), where each domain corresponds to one organization. 
FL offers privacy preserving guarantee in distributed scenarios while DG ensures that the developed model can generalize well to unseen data.
The use of FL in DG is more challenging in practice since the data is inaccessible in this setting. Hence, the assessment of the type of domain shift is harder in this case. The collected data may differ in terms of data acquisition systems, demographics, medical conditions, and treatment protocols. Researcher should choose the most appropriate DG strategy depending on the assumed domain shift (covariate shift or concept shift). 

Notwithstanding this difficulty, federated domain generalization \citep{li2023federated} is a promising research area and one perspective would be to extend more DG algorithms in this context.
One line of research is to use FL mechanisms for improving DG. For instance, \citet{matta2023federated} used FL to study two main factors which affect DL generalizability: the difference
in terms of collected imaging data (screening centers) and the difference of annotation between readers (graders). The targeted application was the detection of diabetic retinopathy using fundus photographs. To this end, they have developed two FL algorithms: 1) a cross-center FL algorithm, using data distributed across centers and 2) a cross-grader FL algorithm, using data distributed across the graders. The study has shown that the cross-grader FL algorithm has outperformed the cross-center FL algorithm and centralized learning (a learning paradigm where all data is pooled in a centralized repository). It suggests that the averaging mechanism used in FL allows to give equal weight to all graders, leading to a more generalized model.

\subsubsection{Fairness}
\label{sec: Fairness and covariate shift}
In the context of covariate shift, some work have proposed to study the performance by attributes. This permits to gain a better understanding of the model biases caused by different dataset shift. For instance, in the mammography field, mass detection performances were analyzed according to mass status, mass size, age, and breast density \citep{garrucho2022domain}. 
Using this analysis, the authors observed that the model seems to have a bias towards masses smaller than five millimeters in diameter and bounding boxes with a high height-to-with ratio, possibly because these samples were not represented in the training dataset.

In histopathology, \citet{graham2023screening} proposed to develop a DL algorithm to screen for colon cancer based on WSI. To investigate potential biases and ensure fairness in the model's predictions, the authors assessed model performances across different demographic subgroups, including sex, age, ethnicity and anatomical site of the biopsy. The differences in model performance based on sex and ethnicity are minimal, but the impact of age on performance is more significant. This variation could stem from several sources, including the data selected for training the model and possible differences in how diseases manifest across different age groups.

In radiology and dermatology, \citet{brown2023detecting} investigated unfairness of DL models due to shortcut learning, a phenomenon where DL models make predictions based on incorrect correlations found in the training data. 
In their experiment on X-ray dataset, the authors have shown that the performances of the DL models varies with age. In addition, these models learn to encode age even though the models were  trained to do so. To identify the presence of shortcut learning when attributes might be causally related to the outcome (such as age), they proposed ShorT, an approach based on adversarial learning. It applies an intervention that modifies the amount of age encoding in the feature extractor and assess the effect of this intervention on model fairness. 

\subsubsection{Causality}

Causal machine learning \citep{kaddour2022causal} is a learning paradigm that utilizes causal knowledge about the to-be-modeled system.
Essentially, causal inference offers a framework for formalizing structural knowledge about the data generating process via Structural Causal Models (SCMs). SCMs permit to estimate the impact on data when changes (called interventions) are applied to its generating process. Moreover, they also allow us to model the consequences of changes in hindsight while taking into account what happened (called counterfactuals). 

One of the most promising areas where causal machine learning can be applied is  DG. Causality aware DG aims to reduce dependency on spurious correlations by addressing and adjusting for confounding variables \citep{sheth2022domain}.
For additional information about these methods, readers are invited to refer to the survey on DG and causality available in the literature, in \citet{sheth2022domain,sheth2023causal} and a survey of causality in medical image analysis \citep{vlontzos2022review}. 

\subsection{What are the best practices for implementing generalization techniques in research?}
In this section, we present generalization libraries, model selection strategies, and evaluation research, for the purpose of readers intending to start employing DG approaches.
\label{sec:Implementation}
\subsubsection{Generalization libraries}
A few general-purpose libraries exist for covariate shift, concept shift and noisy label management. These libraries implement multiple algorithms and benchmarking mechanisms and can therefore be useful to develop DG approaches.
\begin{itemize}
    \item \textbf{DomainBed}\footnote{\href{https://github.com/facebookresearch/DomainBed} {https://github.com/facebookresearch/DomainBed}}\citep{gulrajani2020search}, a testbed for domain generalization, is a PyTorch suite containing benchmark datasets (mainly computer vision datasets) and algorithms for DG. Initially, it includes seven multi-domain datasets, nine baseline algorithms, and three model selection criteria. 
    \item \textbf{Cleanlab}\footnote{\href{https://github.com/cleanlab/cleanlab}{https://github.com/cleanlab/cleanlab}} is a popular library for noisy label management. It implements various data-centric AI algorithms, in which noisy labels are ``cleaned'' before training. Benchmarking relies on a noise generation module.
    \item \textbf{DeepDG}\footnote{\href{https://github.com/jindongwang/transferlearning/tree/master/code/DeepDG}{https://github.com/jindongwang/transferlearning/tree/master/code/DeepDG}}, inspired by DomainBed, DeepDG is a PyTorch based toolkit for DG. It is a simplified version of DomainBed while it adds new features to enhance functionality.
    \item \textbf{Dassl}\footnote{\href{https://github.com/KaiyangZhou/Dassl.pytorch}{https://github.com/KaiyangZhou/Dassl.pytorch}} \citep{zhou2021domain} is a PyTorch toolbox  developed to support research in domain adaptation and generalization. It comprises methods for single-source domain adaptation, multi-source domain adaptation, domain generalization and semi-supervised learning.
    \item\textbf{ClinicalDG}\footnote{\href{https://github.com/MLforHealth/ClinicalDG}{https://github.com/MLforHealth/ClinicalDG}} A Modified version of DomainBed framework. 

\end{itemize}

\subsubsection{Model selection}
Following \citet{gulrajani2020search}, two potential selection methods can be used as model selection policy, \textit{Training-domain validation set} and \textit{Leave-one-domain-out cross-validation}:
\begin{itemize}
    \item \textbf{Training-domain validation set} consists of splitting the data for each source domain into a training subset and a validation subset. The validation subsets are pooled across all source domains to form an overall validation set. Finally, the model maximizing the score performance on the overall validation set is selected. 
    \item  \textbf{Leave-one-domain-out cross-validation} This strategy assumes the presence of at least two source domains. Therefore, it is applicable in multi-source DG. It consists of leaving one source domain for the validation while using the others for training.
\end{itemize}

\subsubsection{Evaluation}
Some studies have focused on the evaluation of the key driver of covariate shift such as the effect of the physical generation process, i.e., Physical Imaging Parameters (PIPs), on model generalization~\citep{kilim2022physical}. Regarding the concept shift, evaluating the models when the test data contains noisy labels has gained interest in the last few years. For instance, \citet{lovchinsky2019discrepancy} tackled this problem and proposed the \textit{discrepancy ratio} as an evaluation metric. In this section, we present common metrics used for evaluation in the generalization research.

{\textbf{F1 score}}
F1 score was used as a metric for evaluating mitotic figure detections in the MIDOG challenge  \citep{aubreville2023mitosis}. Overall $F_1$ is computed as follow using the counting of all True Postives (TP), False Positives (FP) and False Negatives (FN) detections on slide $i$ for all the $k$ processed slides: 
\begin{equation}
    F_1= \frac {2\sum_i^k TP_i}{2 \sum_i^k TP_i+\sum_i^n FP_i+\sum_i^k FN_i}
\end{equation}

{\textbf{Quadratic Cohen’s Kappa}}
This metric \citep{cohen1960coefficient} compares the performance of the algorithms with the reference standard. It reflects the degree of disagreement, in such a way that more emphasis is given to bigger differences among ratings than to minor differences \citep{sim2005kappa}. It is suitable for multi-category ordinal classification. It was used to assess the algorithms in the PANDA challenge \citep{bulten2022artificial}.

{\textbf{Balanced accuracy}} It is defined as the average of recall obtained on each class. This metric was used for assessing the algorithms performances in the ISIC challenge.

{\textbf{\textit{A}-distance}} A-distance measures the distribution discrepancy\citep{ben2010theory}. The smaller the A-distance, the more domain-invariant the features are. Therefore, it is an indicator of how efficient a method is to reduce cross-domain divergence \citep{chen2020cross}.

{\textbf{Representation shift}}
The \textit{representation shift} (R) is used to quantify the statistical difference between the datasets in the evaluation of DG methods \citep{bayasi2022boosternet, stacke2020measuring}. It computes  the differences in the distribution of layer activations of a model between datasets from two domains, capturing the model perceived similarity between the two datasets. The distributions between the two dataset are likely to be similar (small distances) if the model had learnt domain-invariant features.

\subsection{What are the key challenges and future promises for generalization research?}
\label{sec:Future challenges}
In this section, we discuss future possibilities for generalization research in medical imaging. We include perspectives for exploring important research area related to DG including datasets, modelling pipeline strategies, subpopulation shift, open DG, continual DG, unified benchmarking, privacy concerns and multimodal DG.

\subsubsection{Datasets in DG}
Regarding DG datasets, \citet{kilim2022physical} encouraged to include medical image generation metadata in open source datasets. The goal of using metadata measured with standard international units is to establish a universal standard between distributions generated across the world for all current and future imaging modalities. In addition, future work can use these meta-data describing the generative process of an image in unsupervised and self-supervised algorithms. Also, leveraging such metadata to develop models that are agnostic to physical imaging parameters would be an interesting future direction towards more robust models. 
Indeed, these metadata could be used as a tool for predicting the worst case generalization scenario. 

In comparison to multi-DG, where the information related to domains is needed, single-DG is more easy to tackle in practice since it only requires one single source dataset. In this scenario, it is easier for industries to obtain the rights to access this data. Moreover, the problem of missing domain information (i.e., data's originating center) could be solved using single-DG algorithms. 

For a safe deployment, AI systems in health undergo thorough evaluations for validation purposes. In general, it is assumed that the ground truth is fixed (certain). However, in healthcare, the ground truth may be uncertain. Standard evaluations of AI models often overlook this aspect, which can lead to serious repercussions, such as an overestimation of the models' future performance \citep{gordon2021disagreement}. This is particularly concerning in the medical field, because a lack of robustness may translate into patient risk.

\subsubsection{Modelling pipeline strategies}
\label{sec: Modelling pipeline strategies}
 The majority of the work in generalization research tackled the train-test data shift (also called train-test locus), i.e., considering the classical shift type between train and test data. Other types of shift loci can be investigated as proposed by \citet{hupkes2023taxonomy}. Namely, the fine-tune train test locus which refers to the situation where a model is evaluated on a finetuning test set that has a different distribution from the finetuning training data. In this context, finetuning could be achieved by refining all the model parameters, freezing the network's top layer and training only the dense layers \citep{badjie2022deep}, or a few of the final convolutional layers \citep{diwakaran2023breast}. Another type is the pretrain-train locus which evaluates if a specific pretraining method produces models that are effective when subsequently trained on diverse tasks or domains. This is often evaluated in the case of foundation models.
The pre-train-test locus is encountered when a pre-trained model is tested directly on out-of-domain data.

\subsubsection{Subpopulation shift}
\label{sec: Subpopulation shift}
Biases in DL models, associated with factors such as race, gender or age can result in healthcare disparities and negative patient outcomes. In fact, underrepresented training data can lead to subpotimal DL models. One key contributing factor to this is subpopulation shift, i.e, changes in the proportion of some subpopulations between training and deployment \citep{yang2023change}. In these contexts, DL models may have high overall performance yet still underperform in rare subgroups. Subpopulations shift can be categorized into spurious correlations, attribute imbalance, class imbalance and attribute generalization.
Spurious correlations involve non-causal relationships between the input and the label that may shift during deployment, such as  image backgrounds or texture. Attribute imbalance occurs when certain attributes are sampled with a much smaller probability than others in the training. Class imbalance happens when class labels are distributed unevenly, leading to lower preference for minority labels. Attribute generalization refers to the setting where some attributes are absent in the training domain but present in the testing domain.

\subsubsection{Open DG}
\label{sec: Open-set DG}
In conventional DG, it is assumed that the label space is the same between the source domain and the target domain. However, this assumption does not hold in real applications.  Open DG \citep{shu2021open} addresses the problem of DG when the training and test label spaces are not the same. It is a promising approach to tackle the problem where the label taxonomy is not the same between source datasets.  This problem is often encountered in medical image analysis. For example, for developing a multi-disease AI system, \citet{matta2023towards}
 analyzed the labels of different datasets and converted them into a unified labeling system.
 
A special form of open DG is open-set DG, in which the label space on the source domain is considered a subset of that on the target domain. For instance, \citet{zheng2023single} proposed an open-set single-DG based on multiple cross-matching method. 
Their approach consists in generating auxiliary samples that fall outside the category space of the source domain, thereby enhancing the identification of an unknown class (i.e., class that does not belong to the source domain). Crucially, these produced auxiliary samples do not necessarily align with the novel classes within the target domain.

\subsubsection{Continual DG}
\label{sec: Continual DG}
Conventional DG assumes that multiple source domains are accessible and the domain shift is abrupt. However, this is not universally applicable to all real-world applications
where the data distribution may gradually change over time, especially, in the medical field. In this context, new disease or new biomarkers may arise.
As the domain continues to evolve, new domains will consistently emerge. Re-training DL models, under the conventional scheme of DG, to keep-to-date with both new and existing domains can be both resource-intensive and inefficient.
While the transfer learning paradigm seems to be an effective strategy to solve this problem, it should be carefully applied to DG models. For instance, \citet{garrucho2022domain} demonstrated that fine-tuning a DG model to unseen domain can sometimes decrease performance. In the medical domain, transfer learning faces challenges such as data availability and catastrophic forgetting. Fine-tuning models in new domains can lead to overfitting to less diverse datasets and forgetting previously learned information. This could be attributed to a small dataset or even  noisy data. \citet{samala2020generalization} noted that training with noisy data, even with as few as 10\% corrupted labels, could increase generalization error. Therefore, it is also not recommended to perform transfer learning when the quality of the data is poor.

A potential future direction to address these challenges is continual learning. It permits the model to continuously learn from a sequence of tasks over time while maintaining performances on all experienced tasks. Combining continual learning and DG would enable to model the evolutionary patterns of temporal domains and leveraging these patterns to palliate the distribution shift in the future domains.
Recent work \citep{xie2024evolving} proposed a continual domain generalization over temporal drifts, where the goal is to generalize on new unseen domain given that only data from the current domain is accessible at any given time, while information from past domains is unavailable.

\subsubsection{Unified Benchmarking}
\label{sec: Unified Benchmarking}
From this survey, we can see that there is a variation in the targeted application (histopathology, Xray, fundus photographs, ultrasound). In addition, the training protocol differ from one paper to another (architecture, augmentation strategies, etc) or even in datasets (not the same split was used). This makes the comparison between methods challenging and unfair.  A practical solution for this problem is to organize challenges in domain generalization for medical image classification. This help in ensuring the testing data is the same. However, this strategy does not ensure that the main differences in methods come from other factors such as the backbone used. Therefore, for a better assessment of these methods, there is a need for a unified framework like in DomainBed, or like benchmarking framework used in federated learning for medical field such as Flamby \citep{terrail2022flamby} and MedPerf-FeTS \citep{karargyris2023federated}.

In the last few years, benchmarking has shown a great interest in the medical research community. In general, DG performances are compared to a baseline approach,  \textit{Empirical Risk Minimization} (ERM), where a single model is learned on pooled data across all training sources by minimizing the global average risk. Several applications were targeted, \citet{zhang2021empirical} benchmarked\footnote{\url{https://github.com/MLforHealth/ClinicalDG}} the performance of eight DG methods on multi-site clinical times series from Intensive Care Units (ICUs) and chest X-ray imaging data from four sites.
In line with prior work on general imaging datasets \citep{gulrajani2020search}, their experiments on real-world medical imaging data revealed that the current DG methods do not consistently achieve significant gains in OOD performance over ERM. 
More recently, \citet{che2023towards} targeted DR grading in unseen domains. They presented a unified framework named Generalizable Diabetic Retinopathy Grading Network \footnote{\url{https://github.com/chehx/DGDR}}, which demonstrated promising performances compared to ERM. In addition, for fair evaluations, they have provided a publicly available benchmark, the GDR-Bench Dataset, which includes  eight open-source fundus datasets. In line with this study, future work should aim to propose real-world benchmark datasets for different medical modalities specifically for DG. This initiative would undoubtedly promote standardized evaluation protocols, ensuring consistency and reliability in the assessment of DG methods.

\subsubsection{Privacy concerns}
\label{sec: Privacy concerns}
Learning under domain shift in the medical field is also subject to data privacy and regulatory concerns. In certain cases, it is challenging for a single institution to collect enough diverse data, especially for rare diseases. Multi-DG in these settings can facilitate  data collection from multiple institutions, aiming to develop models that generalize to unseen domains. Integrating federated learning to DG is a promising solution to ensure data privacy and compliance with regulations, enabling collaborative efforts without sharing sensitive patient data directly. This approach not only preserves patient confidentiality but also ensures that collaborative research adheres to legal standards such as data protection laws  GDPR and HIPPA.

\subsubsection{Multimodal DG}
\label{sec: Multimodal DG}
A promising research direction involves integrating medical multimodality into DG. Multimodality encompasses combining various data types such as electronic health records, imaging techniques including 2D and 3D image information \citep{li2024review}, and genomic data. This integration adds complexity to assessing the dataset shift. For instance,  there might be scenarios where data from one modality is missing, some data is noisy, unannotated, has unreliable labels, or is scarce during the training or testing phases. Recent work has targeted to solve these problems. For example, generating missing data using generative models. This, however,  may exacerbate the problem by possibly introducing a generated shift. Despite these challenges, this area of research is crucial as it emulates the comprehensive diagnostic methodology employed by medical professionals, and allows for improved DL performances. Indeed, a future direction involves developing innovative multimodal and multidomain AI models for clinical decision-making using foundation models.

\subsection{Limitations}
\label{sec: Limitations}
One limitation of this work is that we only considered Scopus as a database, which may not be representative of all existing work done in this field. Another limitation is that while we focused on two main shifts, we acknowledge that domain shift is more complex in the medical field. These assumed shift (covariate/concept shifts) assume that one of the probability distribution is fixed. However, in real scenarios, this may also be more complex and both shifts can appear simultaneously. While it is more challenging to tackle both problems, future work handling full shift (covariate shift and concept shift) holds great potential for the clinical world. We note that there is a limited consensus on the terminology used in papers. Shift types are defined differently in some papers and new terms can arise as acquisition shift \citep{garrucho2022domain}. This make the search suboptimal for literature review. A unified terminology as proposed in our work and \citep{moreno2012unifying} would help researchers to rely on  a unifying framework for addressing domain shift.

\section{Conclusion}
\label{sec: Conclusion}
In the medical field, data exhibit different sources of variation: images may be collected from multiple countries and different ethnic group (causing covariate shift), data can be gathered using different criteria (different screening programs), annotations differences, etc. (causing concept shift). To mitigate these challenges, we reviewed state-of-the-art methods for the generalization of DL models in medical image classification and discussed challenges and future research trends for this line of research. We hope that this work will help the research community to tackle the problem of generalization in a variety of applications.
Beyond out-of-domain generalization, achieving a fully trustworthy and responsible model in healthcare requires robustness against  malicious (adversarial) attacks, and interpretability. Securing both the data and the models is crucial, especially in medical diagnosis and clinical settings, given the growing regulatory concerns. Interpretability allows for understanding how a model makes its predictions and assessing their validity, which  builds trust in the model and ensures appropriate use. In addition, for safe deployment in real world clinical applications, AI models must express uncertainty when operating outside their training data range.
Ultimately, the methods discussed in this survey could democratize access to AI by offering a scalable screening, more reliable diagnosis and more equitable access to high quality care. Robust clinical validation across various institutions and demographics would further promote the wider adoption of AI in healthcare.

\section*{Acknowledgement}
This work was supported by the French National Research Agency under the LabCom program (ANR-19-LCV2-0005 - ADMIRE project).

\section*{Declaration of competing interests}
The authors declare that there are no known competing interests that could affect this work.

\section*{Registration and protocol}
The review was not registered. A review protocol was not prepared for this systematic review. 

\bibliographystyle{elsarticle-num-names} 
\bibliography{cas-refs}

\end{document}